\documentclass[a4paper,11pt]{article}
\pdfoutput=1 

\usepackage{jcappub} 
\usepackage{xcolor}
\usepackage[T1]{fontenc} 
\usepackage{bm}
\usepackage{epsfig}
\usepackage[lofdepth,lotdepth,caption=false]{subfig}
\usepackage{graphicx}

\usepackage{float}
\usepackage{amsmath,amssymb}
\usepackage{textcomp}
\usepackage{gensymb}
\usepackage[utf8]{inputenc}
\usepackage{setspace}
\usepackage{hyperref}
\usepackage[normalem]{ulem}
\usepackage{pdflscape}
\usepackage{multirow}
\usepackage{fancyvrb}
\usepackage{cleveref}

\def\beq{\begin{equation}}
\def\enq{\end{equation}}
\def\bea{\begin{eqnarray}}
\def\ena{\end{eqnarray}}

\begin{document}

\title{Limits on the parameter space of (3+2) sterile neutrino scenario by IceCube data}
\author[a]{Emilse Cabrera,}
\author[a]{Arman Esmaili,}
\author[b]{Alexander A. Quiroga}
\affiliation[a]{Departamento de F\'isica, Pontif\'icia Universidade Cat\'olica do Rio de Janeiro,\\
Rua Marquês de São Vicente 225, Rio de Janeiro, Brazil}
\affiliation[b]{ILACVN, Universidade da Integra\c{c}\~{a}o Latino-Americana,\\ Avenida Tacredo Neves 6731, Foz do Igua\c{c}u, Brazil}
\emailAdd{emilsecc@aluno.puc-rio.br}
\emailAdd{arman@puc-rio.br}
\emailAdd{alexander.quiroga@unila.edu.br}
\affiliation

\abstract{The neutrino sector of the standard model of particles can contain more than one sterile neutrino states. Generally, existence of more sterile states leads to better, or at least equally good, fit to the short baseline anomalous data due to the larger number of parameters and interferences which create features in the oscillation pattern. However, for experiments like IceCube, where the sterile states distort the oscillation pattern of high energy atmospheric neutrinos through parametric and MSW resonances, addition of more sterile states leads to a more intense effect. Although the limits on one additional sterile neutrino state by IceCube data have been studied in the literature, bounds on the models with more sterile states are lacking. We analyze the one-year data set of atmospheric neutrinos collected by IceCube during the 2011-2012 and derive the limits on the parameter space of (3+2) scenario with two sterile neutrino states, taking into account the relevant systematic and statistical uncertainties and atmospheric neutrino flux variants. To facilitate the joint analysis of IceCube and short baseline data, we provide the table of $\chi^2$ values from IceCube's data analysis as function of the parameters.}
\maketitle
\date{\today}

\section{\label{sec:intro}Introduction}

The neutrino sector of the standard model of particles, with three active neutrino states, contains seven parameters consisting of three masses, three mixing angles and one CP violating phase (if neutrinos are Majorana particles, two additional phases exist which, however, do not affect the flavor oscillations). A plethora of neutrino oscillation experiments over the last three decades accomplished the measurement of the most of these parameters leaving only the mass scale (which oscillation experiments are not sensitive to), the sign of atmospheric mass-squared difference and the CP violating phase, where the last two will be measured in the near future by the Hyper-Kamiokande~\cite{Hyper-Kamiokande:2018ofw}, DUNE~\cite{DUNE:2020ypp} and JUNO~\cite{JUNO:2015zny} experiments. Although the oscillation data can be perfectly fitted, via a global analysis, by the framework of three active neutrino states and its parameters (see~\cite{Esteban:2020cvm,deSalas:2020pgw,Capozzi:2018ubv}), some anomalies have been shown up hinting at the presence of one or more sterile neutrino states. The LSND experiment~\cite{LSND:2001aii} observed an excess of $\bar{\nu}_e$ events hinting at the $\bar{\nu}_\mu\to\bar{\nu}_e$ oscillation resulting from $\Delta m_{41}^2\approx1~{\rm eV}^2$, where $\Delta m_{41}^2=m_4^2-m_1^2$ is the mass-squared difference between $\nu_4$ (mostly sterile) and $\nu_1$ (mostly active) states. The MiniBooNE experiment~\cite{MiniBooNE:2020pnu} has been designed to test the LSND anomaly and observed somehow similar excesses in both the neutrino and antineutrino channels. Another anomaly, dubbed the {\it reactor antineutrino anomaly}, arose from the re-evaluation of the $\bar{\nu}_e$ flux from reactors~\cite{Mueller:2011nm,Huber:2011wv} which led to $\sim3\%$ increase in the predicted number of events in near (10 m to 1 km distance) detectors with respect to earlier expectations that were matching the measurements in several experiments. This discrepancy has been interpreted by the conversion of a portion of reactor $\bar{\nu}_e$ into sterile neutrino with mass-squared difference in the same ballpark as hinted by LSND and MiniBooNE. Several new experiments have been designed to search for the oscillatory behaviour of active-sterile neutrino oscillation as function of the baseline, the so-called baseline-dependent spectral measurements, such as DANSS~\cite{DANSS:2018fnn}, NEOS~\cite{NEOS:2016wee}, Neutrino-4~\cite{Serebrov:2020kmd}, PROSPECT~\cite{PROSPECT:2020sxr} and STEREO~\cite{STEREO:2019ztb} experiments. Among them, except the Neutrino-4 experiment, no hint on the active-sterile neutrino oscillation has been observed, contradicting severely the reactor neutrino anomaly (see also~\cite{Giunti:2021iti} disputing the claim of Neutrino-4 experiment). Still these experiments are in agreement with another anomaly, dubbed as \textit{gallium anomaly}, originating from the observed deficit in calibration tests of solar neutrino detectors, SAGE~\cite{SAGE:2009eeu} and GALLEX/GNO~\cite{Kaether:2010ag}, by electron-capture sources. Recently, the BEST experiment~\cite{Barinov:2021asz} designed to check this anomaly, did confirmed this deficit which can be interpreted as oscillation of electron neutrino to sterile neutrino with mass-squared difference $\sim\mathcal{O}(1)~{\rm eV}^2$ (see~\cite{Elliott:2023cvh,Giunti:2022btk}). 

Another piece of information about the sterile neutrinos comes from the IceCube experiment. It has been realized that Earth's matter induces parametric~\cite{Liu:1997yb,Liu:1998nb} and MSW~\cite{Nunokawa:2003ep,Yasuda:2000xs} resonances in muon antineutrino channel at energies $\sim\mathcal{O}({\rm TeV})$ if sterile neutrino(s) exist with the mass-squared difference(s) $\sim\mathcal{O}({\rm eV}^2)$. These resonances distort the energy and zenith angle distributions of atmospheric neutrinos at $\gtrsim100$~GeV energies, making the IceCube sensitive to these scenarios. Motivated by these studies~\cite{Choubey:2007ji,Barger:2011rc,Razzaque:2011ab,Razzaque:2012tp,Esmaili:2012nz,Esmaili:2013vza,Esmaili:2013cja}, IceCube collaboration published limits on the $(3+1)$ scenario with one additional sterile neutrino state~\cite{IceCube:2016rnb,IceCube:2020phf} and severely constrained this scenario. However, from the phenomenological point of view, the number of sterile neutrino states can be anything, with the most attractive scenario being three sterile neutrino states matching with the three active neutrino states. For the short baseline experiments, and the associated anomalies, adding more sterile neutrino states increases the number of degrees of freedom which generally leads to a better, or at least equally good, fit to the anomalous data. However, for IceCube experiment the addition of more sterile neutrino states strengthens the resonances, leading to severer bounds on $(3+N)$ scenarios with $N\geq2$. As far as we know, in all the fits to anomalous data with $N\geq2$ sterile neutrino states (see e.g. \cite{Diaz:2019fwt,Cianci:2017okw,Hardin:2022muu,Acero:2022wqg}), the IceCube bounds on the parameter space have not been taken into account, simply because of the lack of this information in the literature. This paper is intended to address this lack. We analyze one-year data of atmospheric neutrinos collected by IceCube and derive the bounds on the parameter space of $N=2$ scenario, that is the $(3+2)$ scenario. We compare this bounds with the allowed regions of $(3+2)$ parameter space from short baseline data fits and find that parts of these regions are excluded by the IceCube data. To make a joint analysis of short baseline experiments and IceCube data feasible, we provide the table of $\chi^2$ values in an ancillary file to this paper. 

The paper is organized as follows. In section~\ref{sec:pheno} we briefly
describe the $(3+2)$ scenario and its parameter space. In section~\ref{sec:parameters} we study the atmospheric neutrino oscillation in $(3+2)$ scenario and single out the relevant parameters for our analysis. The utilized IceCube data set and the analysis method are explained in section~\ref{sec:analysis}. In section~\ref{sec:res} we provide the main results of the analysis, that are the bounds on the parameter space of $(3+2)$ scenario. A summary of the results and conclusions is provided in section~\ref{sec:SumConc}. The numerical method employed for computing the oscillation probabilities of atmospheric neutrinos is summarized in appendix~\ref{sec:Cayley_32}. 

\section{\label{sec:pheno}Neutrino sector with two additional sterile neutrinos: (3+2) scenario}

We consider the extension of neutrino sector in the Standard Model with two additional sterile neutrinos, $\nu_{s_1}$ and $\nu_{s_2}$, totally having five flavor fields $\nu_{\alpha}$ where $\alpha = e, \mu, \tau, s_1, s_2$. The flavor basis states are related to the mass basis states $|\nu_{i}\rangle$, where $i=1,\ldots,5$, by the unitary mixing matrix $U$:
\begin{equation*}
    |\nu_{\alpha}\rangle = \sum_{i}U_{\alpha i}^\ast \,|\nu_{i}\rangle~,
\end{equation*}
where $U_{\alpha i}$'s are the elements of the $5\times5$ unitary mixing matrix (a generalization of the usual $3\times3$ Pontecorvo–Maki–Nakagawa–Sakata matrix), which can be parameterized as 
\begin{equation}
    U = R^{45}(\theta_{45})R^{35}_{\delta_{35}}(\theta_{35})R^{25}_{\delta_{25}}(\theta_{25})R^{15}_{\delta_{15}}(\theta_{15})R^{34}(\theta_{34})R^{24}_{\delta_{24}}(\theta_{24})R^{14}_{\delta_{14}}(\theta_{14})R^{23}(\theta_{23})R^{13}_{\delta_{13}}(\theta_{13})R^{12}(\theta_{12}),
\label{eq:mixing_matrix}
\end{equation}
where $R^{ij}$ ($i,j = 1,\ldots,5$, with adjacent $i$ and $j$ and $i<j$) is the rotation matrix in the $ij$-plane with the angle $\theta_{ij}$. The rotation matrices $R^{ij}_{\delta_{ij}}$ with non-adjacent $i$ and $j$ contain a CP-violating phase $\delta_{ij}$ (the $\delta_{13}$ is the conventional CP-violating phase of the $3\nu$ scheme). The mixing matrix of $(3+2)$ scenario  contains, in addition to the four parameters of $3\nu$ framework, seven more angles and five more CP-violating phases. For future reference, the mixing matrix elements of interest in our study, in the parameterization of Eq.~(\ref{eq:mixing_matrix}), are given by
\begin{equation}\label{eq:Us1}
U_{e4} = s_{14} c_{15} e^{-i\delta_{14}}\quad , \quad U_{e5} = s_{15} e^{-i\delta_{15}}\quad , \quad U_{\mu 5} = s_{25} c_{15} e^{-i\delta_{25}}~, 
\end{equation}
and
\begin{equation}\label{eq:Us2}
   U_{\mu 4} = s_{24} c_{14} c_{25} e^{-i\delta_{24}} - s_{14} s_{15} s_{25} e^{-i(\delta_{14}-\delta_{15}+\delta_{25})}~, 
\end{equation}
where $c_{ij}\equiv \cos \theta_{ij}$ and $s_{ij}\equiv \sin \theta_{ij}$. 

The evolution of neutrino flavor states in the course of propagation in the medium, in our case the Earth, can be found by solving the following Schr\"odinger-like equation
\begin{equation}
i\,\frac{{\rm d} \nu_{\alpha}}{{\rm d}r} = \mathcal{H}_{\alpha\beta} \nu_{\beta}~,  
\label{ODE}
\end{equation}
with the effective Hamiltonian
\begin{equation*}
    \mathcal{H}_{\alpha\beta} = \left[\frac{1}{2 E_{\nu}} \left(UM^{2}U^{\dagger} + A(r)\right)\right]_{\alpha\beta}~,
\end{equation*}
where $E_{\nu}$ is the neutrino energy, $M^{2}$ is the diagonal neutrino mass-squared matrix
\begin{equation}
M^{2} = \text{diag}(0, \Delta m_{21}^{2},\Delta m_{31}^{2},\Delta m_{41}^{2},\Delta m_{51}^{2})~,
\label{mass_dig_matrix}
\end{equation}
and the effective matter potential matrix $A(r)$ is given by
\begin{equation}
A(r) = 2\sqrt{2}\,G_{F}E_{\nu}\,\text{diag}(N_{e}(r),0,0,N_{n}(r)/2,N_{n}(r)/2)~,
\end{equation}
where $G_{F}$ is the Fermi constant, $N_{e}(r)$ and $N_{n}(r)$ are the electron and neutron number densities of the Earth, respectively. In finding the numerical solution of Eq.~(\ref{ODE}), the $N_{e}(r)$ and $N_{n}(r)$ have been extracted from the Preliminary Reference Earth Model (PREM)~\cite{Adam:1981prem} of Earth's density profile. For the evolution of antineutrinos, the equations are similar to those used for neutrinos in Eq.~\eqref{ODE}, with the replacement $U \rightarrow U^{*}$ and $A(r)\rightarrow -A(r)$.

The oscillation probabilities, for all the energies and zenith angles of interest and for neutrinos/antineutrinos, have been found by numerically solving Eq.~\eqref{ODE} with the Cayley-Hamilton method. The details of this method are described in appendix~\ref{sec:Cayley_32}.

\section{\label{sec:parameters}Atmospheric neutrino oscillations in (3+2) scenario}

The $(3+2)$ scenario has 20 parameters: 10 mixing angles, 6 CP-violating phases and 4 mass-squared differences; which clearly is not amenable to any statistical analysis. However, not all of these parameters are needed to be considered. At high energies, $\gtrsim100$~GeV, no standard oscillation of flavors occur for neutrinos crossing the Earth: for $\nu_e$ the mixing angle is strongly suppressed and is practically zero at these energies; while for $\nu_\mu$ and $\nu_\tau$ flavors, although the mixing angle is unsuppressed, the oscillation length increases with energy and becomes much bigger than the diameter of the Earth. Thus, the five parameters of standard oscillation can be neglected. Also, the angle $\theta_{45}$, quantifying the mixing between the two sterile states, does not enter any physical quantity since it can always be rotated away by a redefinition of the sterile states. 

The flux of atmospheric neutrinos at high energies, $\gtrsim100$~GeV, mainly consists of muon neutrinos and antineutrinos, with electron flavor neutrino flux at least $\sim1/20$ smaller~\cite{Honda:2006qj}, which can be safely neglected for our purposes. Since the analysis we are performing treats the muon-track events collected by IceCube, the two oscillation probabilities of interest are $P(\nu_\mu\to\nu_\mu)$ and $P(\bar{\nu}_\mu\to\bar{\nu}_\mu)$. It is well-known that the existence of sterile neutrinos leads to MSW~\cite{Nunokawa:2003ep,Yasuda:2000xs} and parametric~\cite{Liu:1997yb,Liu:1998nb} resonances for muon antineutrinos traversing the Earth's matter at energies $E_\nu\sim4\,{\rm TeV} \left( \Delta m^2/{\rm eV}^2\right)$ and $E_\nu\sim2.3\,{\rm TeV} \left( \Delta m^2/{\rm eV}^2\right)$, respectively, where $\Delta m^2$ is the mass-squared difference between the (mostly) active and sterile states. The mixing parameters that control these resonances are $\theta_{24}$, $\theta_{25}$, $\theta_{34}$, $\theta_{35}$, $\delta_{24}$, $\delta_{25}$ and $\delta_{35}$. Thus, the probabilities $P(\nu_\mu\to\nu_\mu)$ and $P(\bar{\nu}_\mu\to\bar{\nu}_\mu)$ are independent of $\theta_{14}$, $\theta_{15}$, $\delta_{14}$, $\delta_{15}$ and we can set them to zero in the analysis of IceCube data. For short baseline experiments, such as LSND and MiniBooNE, the oscillation depth (in the appearance channels $\nu_e(\bar{\nu}_e)\to\nu_\mu(\bar{\nu}_\mu)$) depends on the following combinations of elements: $|U_{e4}|^2|U_{\mu4}|^2$, $|U_{e5}|^2|U_{\mu5}|^2$ and $|U_{e4}||U_{\mu4}||U_{e5}||U_{\mu5}|$ and ${\rm arg}\{ U_{e5}^\ast U_{\mu5} U_{e4} U_{\mu4}^\ast \}$; which means dependence on the mixing angles $\theta_{14}$, $\theta_{15}$, $\theta_{24}$, $\theta_{25}$, and the combination of CP-violating phases $\delta_{14}-\delta_{15}+\delta_{25}-\delta_{24}$.   

\begin{figure}[t]
    \centering
    \includegraphics[width=0.98\textwidth]{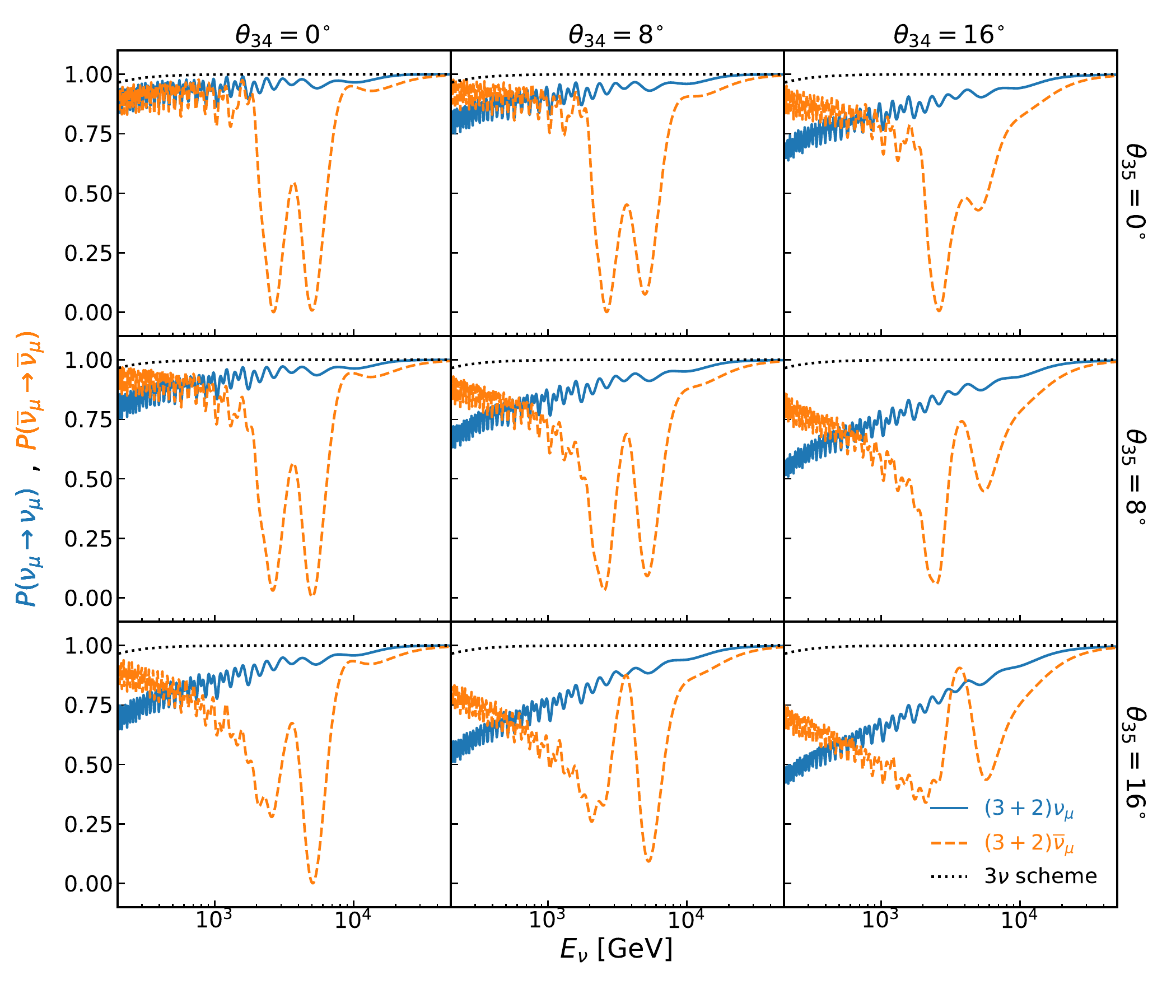}
    \caption{Survival probabilities of core-crossing ($\cos\theta_z=-1$) atmospheric $\nu_{\mu}$ (blue solid) and $\bar{\nu}_{\mu}$ (orange dashed) neutrinos, for three different values of $\theta_{34}$ and $\theta_{35}$. All the CP-violating phases are set to zero and we assume $\theta_{24} = \theta_{25} = 8^{\circ}$, $\Delta m_{41}^{2}=1$~eV$^{2}$ and $\Delta m_{51}^{2} = 2$~eV$^{2}$. The black dotted curve shows the oscillation probability in $3\nu$ standard framework.}
    \label{fig:t34_t35}
\end{figure}

Still, it seems that nine parameters (four angles, three phases and two mass-squared differences) control the resonances in $\bar{\nu}_\mu\to\bar{\nu}_\mu$ oscillation probability of atmospheric neutrinos in $(3+2)$ scenario; yet far from being manageable by the available computing resources. However, at least for a conservative analysis, not all of these parameters need consideration. The dependence of $\nu_\mu(\bar{\nu}_\mu)\to\nu_\mu(\bar{\nu}_\mu)$ oscillation probabilities on $\theta_{34}$ and $\theta_{35}$ angles are shown in Figure~\ref{fig:t34_t35}, for core-crossing trajectories ($\cos\theta_z=-1$) while we set $\theta_{24}=\theta_{25}=8^\circ$, $\Delta m_{41}^{2} = 1$~eV$^{2}$ and $\Delta m_{51}^{2} = 2$~eV$^{2}$. In the upper-left panel (for $\theta_{34}=\theta_{35}=0$) the parametric resonances at $\sim2.3$~TeV and $\sim4.6$~TeV, respectively corresponding to $\Delta m_{41}^{2}$ and $\Delta m_{51}^{2}$, are evident. As can be seen, by increasing $\theta_{34}$ or $\theta_{35}$, the dips in the $P(\bar{\nu}_\mu\to\bar{\nu}_\mu)$ orange dashed curves become shallower, while at the same time both the $P(\nu_\mu\to\nu_\mu)$ (blue solid curves) and $P(\bar{\nu}_\mu\to\bar{\nu}_\mu)$ (orange dashed curve) diverge more from the dotted black curve, corresponding to the standard $3\nu$ oscillation, at lower energies. Since the atmospheric flux is larger at lower energies, and the muon neutrino flux is larger at least by a factor of few than the muon anti-neutrino flux, the overall effect of increasing the $\theta_{34}$ and $\theta_{35}$ is strengthening the limits on the parameters of $(3+2)$ scenario. The physics behind this strengthening has been studied and justified analytically in~\cite{Esmaili:2013vza} within the $(3+1)$ scenario. The same arguments apply here for the $(3+2)$ scenario. Thus, we can set $\theta_{34}=\theta_{35}=0$ in our analysis since their (unknown) non-vanishing values always lead to stronger bounds on the other parameters. 

\begin{figure}[t]
 \centering
  \subfloat{
   \label{fig:d24_nu}
    \includegraphics[width=0.5\textwidth]{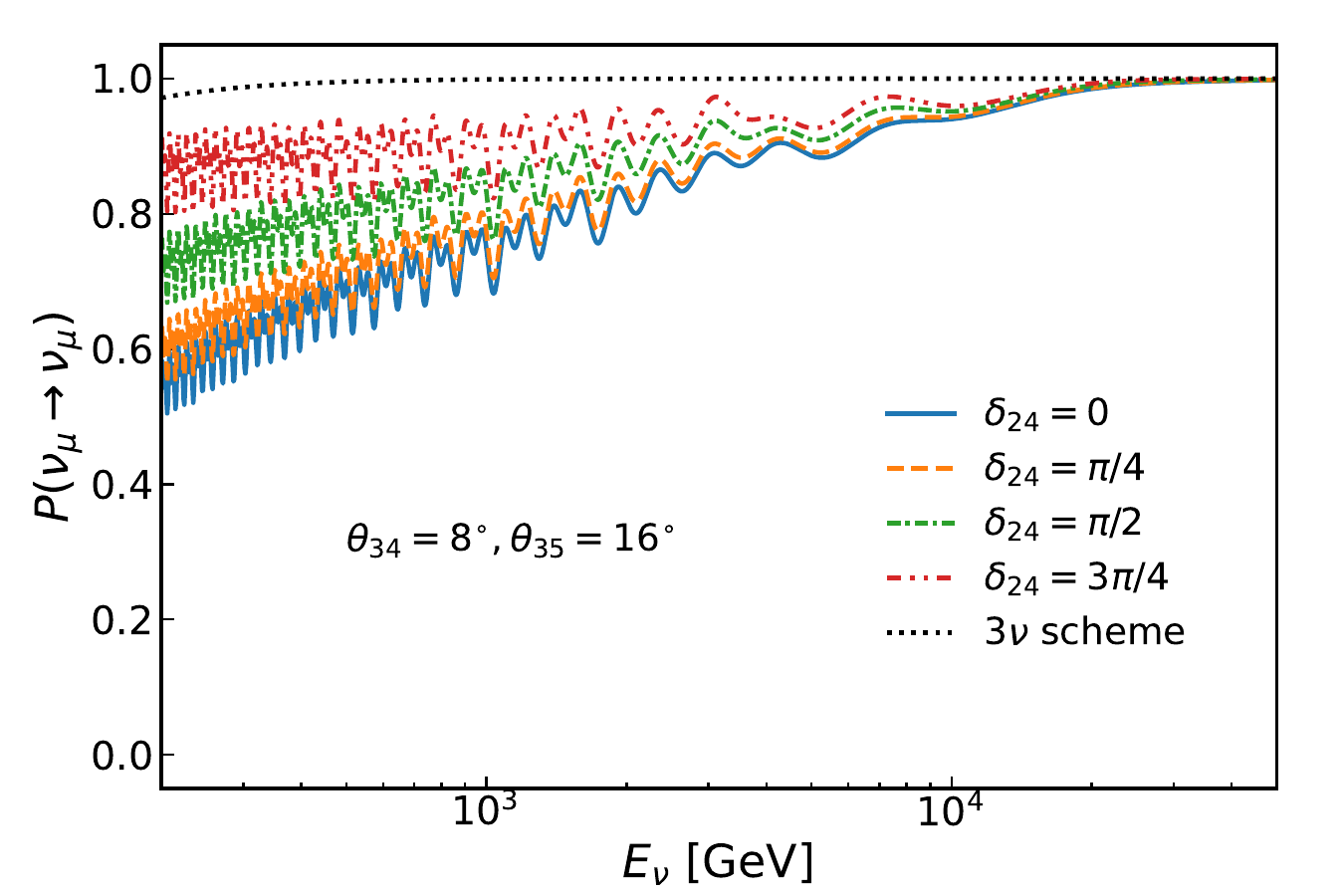}}
  \subfloat{
   \label{fig:d35_nu}
    \includegraphics[width=0.5\textwidth]{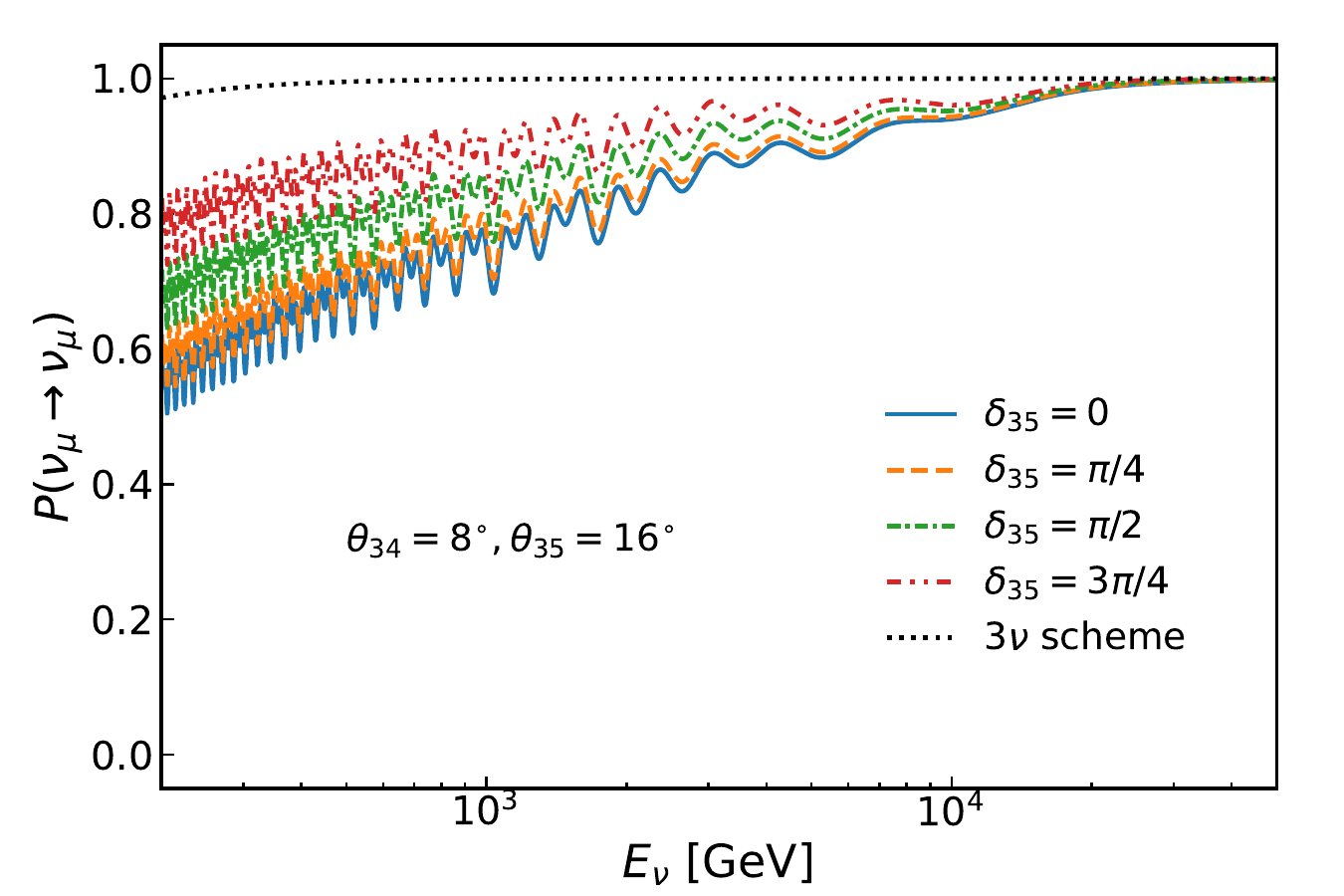}}\\
  \subfloat{
   \label{fig:d24_antinu}
    \includegraphics[width=0.5\textwidth]{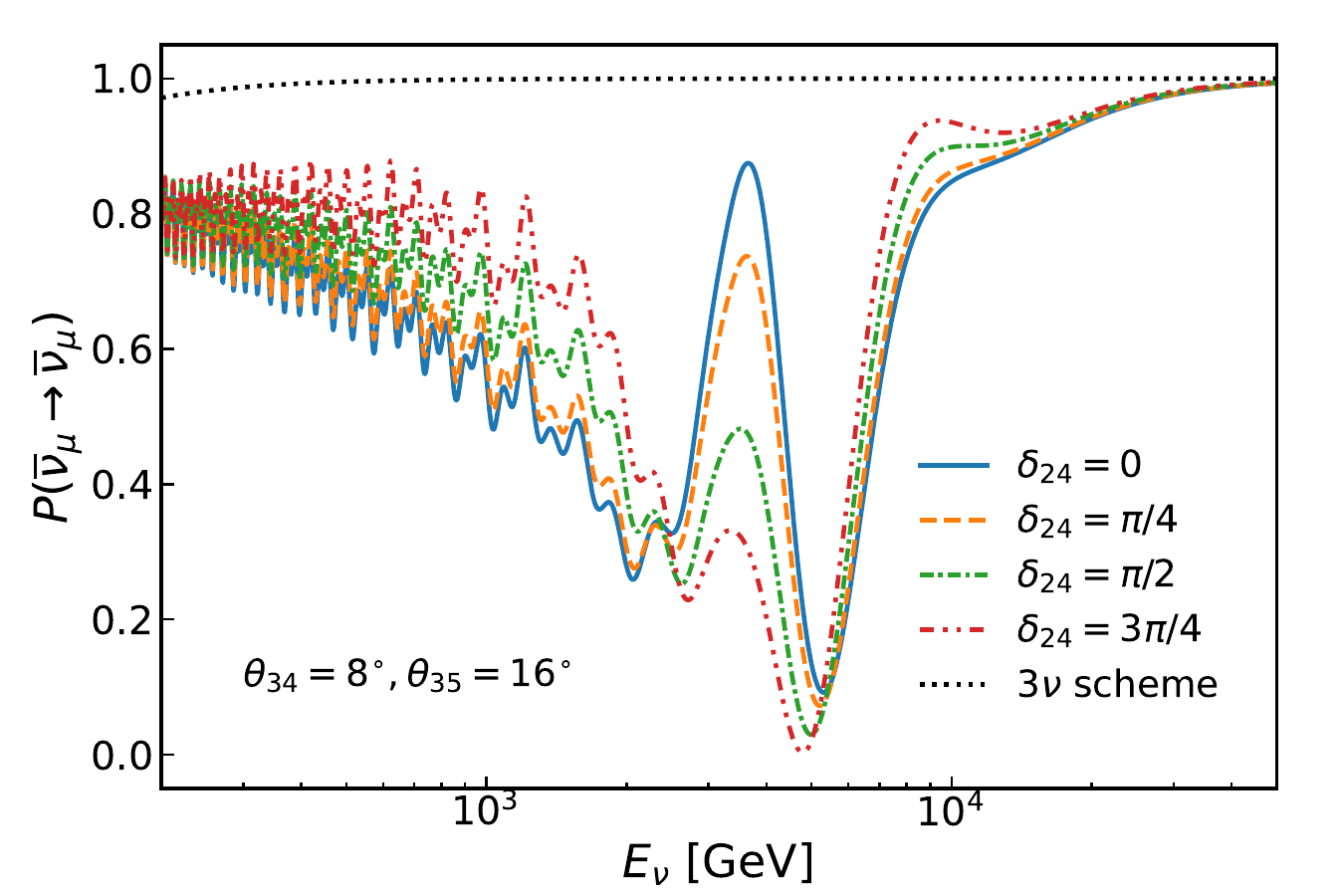}}
  \subfloat{
   \label{fig:d35_antinu}
    \includegraphics[width=0.5\textwidth]{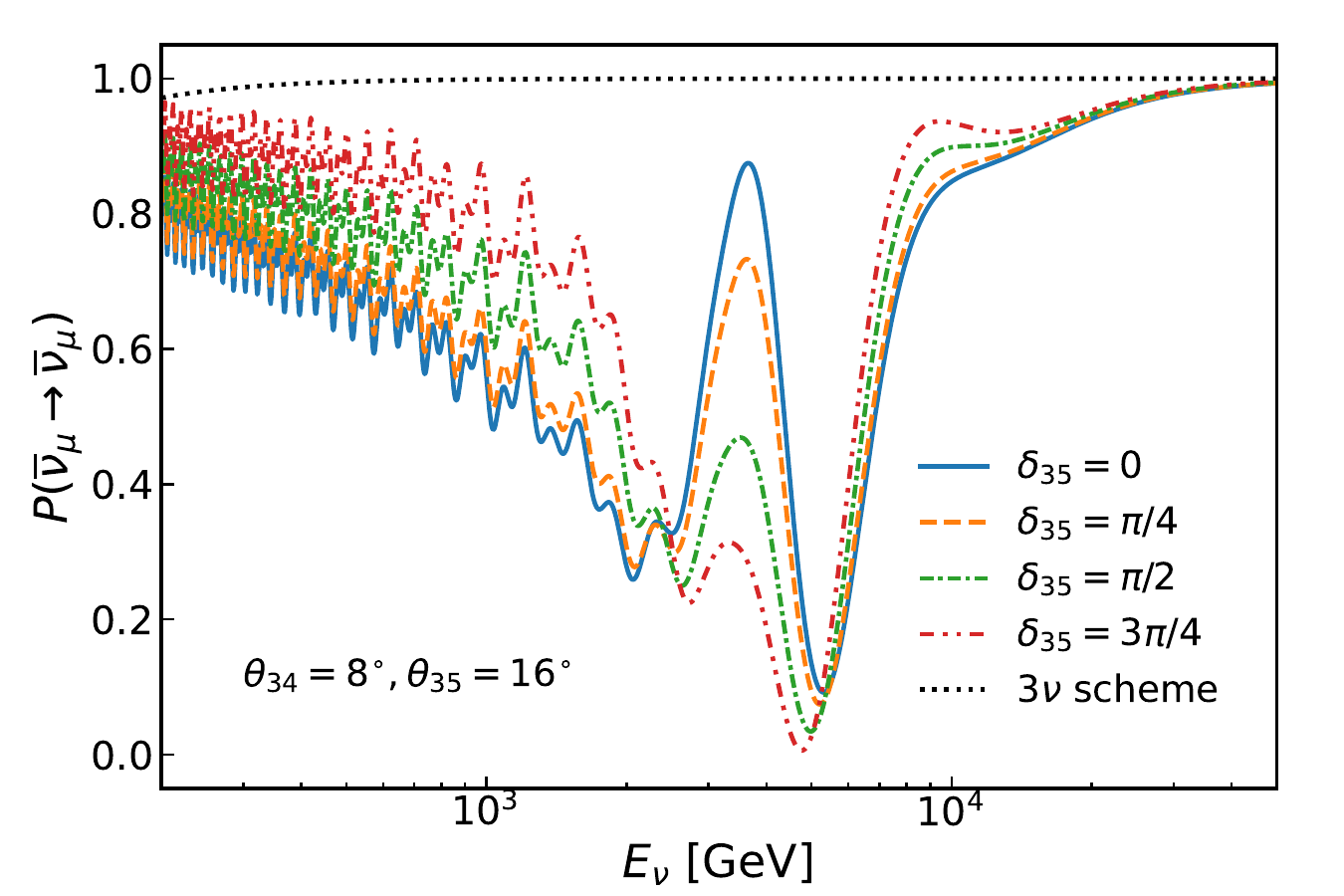}}
 \caption{Dependence of the $P(\nu_\mu\to\nu_\mu)$ (upper row) and $P(\bar{\nu}_\mu\to\bar{\nu}_\mu)$ (lower row) on the CP-violating phases $\delta_{24}$ (left panels) and $\delta_{35}$ (right panels) for core-crossing ($\cos\theta_z=-1$) atmospheric neutrinos. The black dotted curve in all the panels is for standard $3\nu$ oscillation. We assumed $\theta_{24}=\theta_{25}=8^\circ$, $\theta_{34}=8^\circ$, $\theta_{35}=16^\circ$, $\Delta m_{41}^{2}=1$~eV$^{2}$ and $\Delta m_{51}^{2} = 2$~eV$^{2}$.}
 \label{fig:CP_phases}
\end{figure}

Although by setting $\theta_{34}=\theta_{35}=0$, as justified above, the CP-violating phases can be reabsorbed by a redefinition of neutrino fields, whether these phases can relax the limits when $\theta_{34}\neq0$ and/or $\theta_{35}\neq0$ or not, remains to be answered. The dependence of $P(\nu_\mu\to\nu_\mu)$ and $P(\bar{\nu}_\mu\to\bar{\nu}_\mu)$, respectively in the upper and lower rows, on $\delta_{24}$ and $\delta_{35}$, respectively in the left and right panels, are shown in Figure~\ref{fig:CP_phases}. For these figures we have chosen $\cos\theta_z=-1$, $\theta_{24}=\theta_{25}=8^\circ$, $\theta_{34}=8^\circ$, $\theta_{35}=16^\circ$, $\Delta m_{41}^{2}=1$~eV$^{2}$ and $\Delta m_{51}^{2} = 2$~eV$^{2}$. Nonzero values of $\delta_{24}$ and $\delta_{35}$ makes the $P(\nu_\mu\to\nu_\mu)$ closer to $3\nu$ oscillation probability at lower energies, though not completely equal. However, in the muon antineutrino channel, the nonvanishing phases lead to a widening of resonance region while the depth of resonance dips barely change. This widening results in stronger bounds for nonvanishing CP-violating phases $\delta_{24}$ and $\delta_{35}$. The dependence of probabilities on $\delta_{25}$ is very similar to the dependence on $\delta_{24}$ shown in Figure~\ref{fig:CP_phases}. For a more comprehensive discussion of the effect of CP-violating phase on the oscillation probability we refer the readers to~\cite{Esmaili:2013vza}. In the following we will set $\delta_{24}=\delta_{25}=\delta_{35}=0$ to obtain robust and conservative bounds on the other parameters, that are $(\theta_{24},\theta_{25},\Delta m_{41}^{2},\Delta m_{51}^{2})$.  

To summarize, we conclude from the discussions in this section that for the purpose of probing the $(3+2)$ scenario by the IceCube atmospheric neutrino data, it is sufficient to keep only the $(\theta_{24},\theta_{25},\Delta m_{41}^{2},\Delta m_{51}^{2})$ parameters in the analysis and set the rest to zero. Nonzero values of the parameters $(\theta_{34},\theta_{35},\delta_{24},\delta_{25},\delta_{35})$ lead to stronger limits on the parameter space of $(3+2)$ scenario and, following a conservative approach, we will ignore them in our analysis.

\section{\label{sec:analysis}IceCube data set and analysis method}

We compute the survival probabilities of atmospheric $\nu_\mu$ and $\bar{\nu}_\mu$ numerically, as function of $E_\nu$ and $\cos\theta_z$ and scanning over the parameter space of $(\theta_{24},\theta_{25},\Delta m_{41}^{2},\Delta m_{51}^{2})$. To derive the bounds on the parameter space we use the public data of IceCube collected over a period of 343.7 days between 2011-2012\footnote{Available at \url{https://icecube.wisc.edu/science/data-releases/}} which has been used for probing the $(3+1)$ scenario~\cite{IceCube:2016rnb}. This data set contains 20,145 well-reconstructed muon-track events with the reconstructed muon energies (a proxy of neutrino energy providing a lower limit on it) in the range $0.4 \leq E_{\mu}/\text{TeV} \leq 20$ and the reconstructed zenith angles between $-1 \leq \cos \theta_{z} \leq 0.24$. Since in the high energy the difference between the zenith angle of the incoming neutrino and the produced muon is negligible (less than $\sim 0.5^\circ$ at $E_\nu =10$~TeV), we assume that the two are equal. However, the muon energy proxy, $E_\mu$, serves only as a lower limit on the true energy, $E_\nu$, of the incoming neutrino. The relation between them, which depends also on the optical efficiency of detector and incoming direction, is provided by the IceCube collaboration via the \textit{response array} tensor for neutrinos, $\eta_{ijk}$, and antineutrinos, $\bar{\eta}_{ijk}$, where the $i$, $j$ and $k$ indices label the binnings of $E_\nu$, $E_\mu$ and $\cos\theta_z$, respectively. In our analysis, we restrict the neutrino energy to the range $0.2\leq E_{\nu}/\text{TeV} \leq 10^{3}$ and divide the range into 200 bins. For the zenith angle we use the range $-1 \leq \cos \theta_{z} \leq 0$ divided into 17 bins; while the muon energy's range is $0.4 \leq E_{\mu}/\text{TeV} \leq 20$ divided into 10 bins. All the binning patterns follow the published data set by IceCube~\cite{IceCube:2016rnb}.  

The expected number of events, $N_{jk}^{\text{exp}}$, in the given $(j,k)$th bin of observable reconstructed-quantities, $E_{\mu}$ and $\cos\theta_{z}$, is given by 
\begin{equation}\label{eq:N_exp}
N_{jk}^{\text{exp}}=\sum_{i} \left[\eta_{ijk} \phi^{\nu,{\rm atm}}_{ik} \langle P(\nu_{\mu}\rightarrow \nu_{\mu})\rangle_{ik} + \bar{\eta}_{ijk} \phi^{\bar{\nu},{\rm atm}}_{ik} \langle P(\bar{\nu}_\mu\rightarrow \bar{\nu}_\mu)\rangle_{ik} \right]~,    
\end{equation}
where $\langle P(\nu_\mu\to\nu_\mu)\rangle_{ik}$ and $\langle P(\bar{\nu}_\mu\to\bar{\nu}_\mu)\rangle_{ik}$ are respectively the averaged survival probability of $\nu_\mu$ and $\bar{\nu}_\mu$ in the $(i,k)$th bin of $E_{\nu}$ and $\cos\theta_{z}$. The $\phi^{\nu,{\rm atm}}_{ik}$ in Eq.~(\ref{eq:N_exp}) denotes the average atmospheric $\nu_\mu$ flux in the $(i,k)$th bin originating from the decay of pions, $\phi^{\nu,\pi}_{ik}$, and kaons, $\phi^{\nu,K}_{ik}$, which are produced by the cosmic ray interactions with the atmosphere, and is given by 
\begin{equation}\label{eq:nuflux}
\phi^{\nu,{\rm atm}}_{ik}= N_{0}\left[\phi^{\nu,\pi}_{ik} + \left(1.1-R_{\pi/K}\right) \phi^{\nu,K}_{ik} \right] \left(\frac{E_{\nu,i}}{E_{0}}\right)^{-\gamma}~,
\end{equation}
where $N_0$, $R_{\pi/K}$ and $\gamma$ are the parameters that take into account the uncertainties in the normalization, the pion/kaon ratio and energy index of the flux, respectively. The $E_0=13.7$~TeV is the median of the total energy range considered in the analysis and $E_{\nu,i}$ is the median energy in the $i$'th bin of neutrino energy. The flux of atmospheric muon antineutrinos is given by 
\begin{equation}\label{eq:anuflux}
 \phi^{\bar{\nu},{\rm atm}}_{ik} = N^\prime N_{0}\left[\phi^{\bar{\nu},\pi}_{ik} + \left(1.1-R_{\pi/K}\right) \phi^{\bar{\nu},K}_{ik} \right] \left(\frac{E_{\nu,i}}{E_{0}}\right)^{-\gamma}~,   
\end{equation} 
where the $N^\prime$ parameterizes the uncertainty in the relative normalization of neutrino and antineutrino fluxes. 

The quantities $\phi^{\nu,\pi}_{ik}$, $\phi^{\nu,K}_{ik}$, $\phi^{\bar{\nu},\pi}_{ik}$ and $\phi^{\bar{\nu},K}_{ik}$ are provided by the IceCube for seven different models of the atmospheric neutrino flux consisting of the HKKM model~\cite{Honda:2006qj,Sanuki:2006yd,Gaisser:2013bla} and combinations of three primary cosmic ray models (Gaisser-Hillas~\cite{Gaisser:2013bla}, Polygonato~\cite{Hoerandel:2002yg} and Zatsepin-Sokolskaya~\cite{Zatsepin:2006ci}) with two different hadronic models (QGSJET-II-4~\cite{Ostapchenko:2010vb} and SIBYLL2.3~\cite{Riehn:2015oba}). 

The limits on the parameter space of $(3+2)$ scenario are derived by a binned Poisson log-likelihood analysis with added pull-terms for the continuous nuisance parameters $\vec{\xi}=(N^\prime,N_0,R_{\pi/K},\gamma)$. Discrete nuisance parameters consist of the seven atmospheric neutrino flux models and nine variants of response array tensor: four different digital optical module efficiencies, four models of the ice and a nominal response array of the detector. Marginalizations have been performed over the $\vec{\xi}$ and the following combination of flux and detector variants: taking the Honda-Gaisser flux we marginalize over the response array variants and taking the nominal response array we marginalize over the flux models. The $\chi^2$ function is given by
\begin{align}
            \chi^2(\vec{\theta}) &=-2\ln \mathcal{L} (\vec{\theta})  = & \\ 
            & \text{min}_{\Vec{\xi},\lbrace d \rbrace}\left( 2\sum_{j,k}\left[ N_{jk}^{\text{exp}}(\Vec{\theta},\Vec{\xi},d)- N_{jk}^{\text{obs}} + N_{jk}^{\text{obs}} \ln \frac{N_{jk}^{\text{obs}}}{N_{jk}^{\text{exp}}(\Vec{\theta},\Vec{\xi},d)}\right] + \sum_{\eta} \frac{\left(\xi_{\eta}-\Xi_{\eta}\right)^{2}}{\sigma_{\eta}^{2}}\right)~, \nonumber &
\end{align}
where $\Vec{\theta}=(\Delta m_{41}^2,\Delta m_{51}^2,\sin^22\theta_{24},\sin^22\theta_{25})$ is the set of parameters in $(3+2)$ scenario and the set of discrete nuisance parameters is denoted by $\lbrace d \rbrace = \lbrace \text{flux models, detector variants}\rbrace$. The $N_{jk}^{\text{obs}}$ is the observed number of events in the $(j,k)$th bin. The $\eta$ index runs over the continuous nuisance parameters, where the $\Xi_\eta$ and $\sigma_\eta$ are the central values and the $1\sigma$ Gaussian widths of these parameters reported in Table~\ref{tab:nuisance}.  

\begin{table}[H]
    \centering
    \caption{The continuous nuisance parameters, their central values and the $1\sigma$ Gaussian widths.}
    \begin{tabular}{|c|c|c|}
    \hline 
    Parameter &  Central value & Gaussian width\\ \hline\hline
    Normalization ($N_0$) & 1 & 0.4 \\ \hline
    $\nu / \bar{\nu}$ normalization ($N^\prime$) & 1 & 0.025 \\ \hline
    $\pi / K$ ratio ($R_{\pi/K}$) & 0 & 0.1 \\ \hline
    Energy index ($\gamma$) & 0 & 0.05\\ \hline
    \end{tabular}
    \label{tab:nuisance}
\end{table}

The $\chi^2(\vec{\theta})$ has been calculated for $\Delta m_{41}^2$ and $\Delta m_{51}^2$ in the range $[0.1,16]$~eV$^2$, and $\sin^22\theta_{24}$ and $\sin^22\theta_{25}$ in the range $[0.01,1]$. From this $\chi^2$ function, 2-dimensional bounds on the parameters can be derived via the standard procedure of marginalization over the rest of the parameters, which are presented in the next section. To facilitate the reproduction of our analysis and incorporation in a joint analysis with other experiments, such as short baseline experiments, we provide the table of $\chi^2$ values as function of $(\Delta m_{41}^2,\Delta m_{51}^2,\sin^22\theta_{24},\sin^22\theta_{25})$, after the marginalizations over nuisance parameters, as an ancillary file to this manuscript.

\section{Results\label{sec:res}}

From the $\chi^2(\vec{\theta})$ we can find the allowed region of the $(3+2)$ parameter space by the IceCube data. The best-fit point in our analysis is $\Delta m_{41}^2 = \Delta m_{51}^2 =16$ eV$^2$, $\sin^22\theta_{24}=0.3$ and $\sin^22\theta_{25}=0.23$. The value of $\chi^2$ at the best-fit point is $\chi^2_{\rm min}= 179.52$, with the reduced value $\chi^2_{\rm min}/\text{dof} = 1.08$, where dof is the number of degrees of freedom equal to the number of bins, 170, minus the number of parameters. The reduced value 1.08 of $\chi^2$ indicated an overall good fit to the data. The large values of the best-fits of $\Delta m_{41}^2 $ and $\Delta m_{51}^2$, which effectively do not introduce any distortion to the oscillation pattern in the energy range of data, indicates that the IceCube data do not show any preference to $(3+2)$ scenario over the standard $3\nu$ scheme. For the $3\nu$ scenario we obtain $\chi^2(\vec{\theta}=0)= 181.78$, which means that the data only marginally prefer the obtained best-fit in the $(3+2)$ scenario, but not at any significant confidence level.

\begin{figure}[t]
    \centering
    \includegraphics[width=0.75\textwidth]{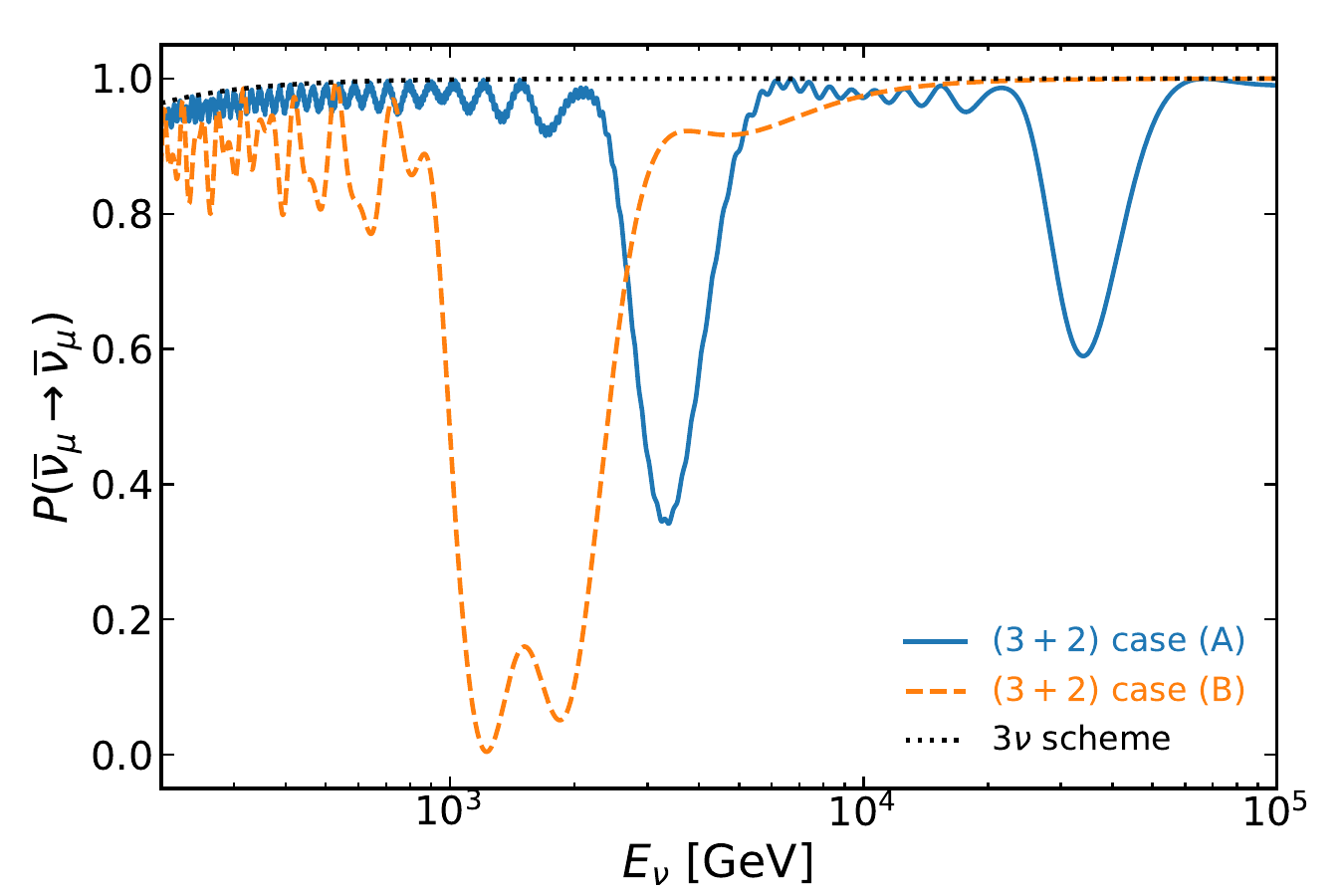}
    \caption{Survival probability of $\bar{\nu}_{\mu}$ as a function of neutrino energy for core-crossing trajectory ($\cos\theta_z=-1$) in the $3\nu$ scheme (black dotted curve) and in the $(3+2)$ scenario for two different best-fits from short baseline experiments, denoted as cases (\textbf{A}) and (\textbf{B}), taken respectively from \cite{Diaz:2019fwt} and \cite{Cianci:2017okw} (see Table~\ref{tab:fits}).}
    \label{fig:prob-cases}
\end{figure}

From the available fits to the \textit{anomalies} in the short baseline experiments, we choose two cases from \cite{Diaz:2019fwt} and \cite{Cianci:2017okw}, respectively denoted as case (\textbf{A}) and (\textbf{B}). Natural parameters for the fits to the short baseline data are the matrix elements $|U_{e4}|$, $|U_{e5}|$, $|U_{\mu4}|$, $|U_{\mu5}|$, and a CP-violating phase usually denoted by $\phi_{54}$, equal to $\delta_{14}-\delta_{15}+\delta_{25}-\delta_{24}$ in our parameterization, in addition to the two mass-squared differences in $(3+2)$ scenario. The values of the parameters in our parameterization of mixing matrix in Eq.~(\ref{eq:mixing_matrix}), corresponding to the best-fit points obtained in cases (\textbf{A}) and (\textbf{B}), can be found via the relations in Eqs.~(\ref{eq:Us1}) and (\ref{eq:Us2}), leading to the reported values in Table~\ref{tab:fits}. To illustrate these cases, in figure~\ref{fig:prob-cases} we show the oscillation probability $P(\bar{\nu}_\mu\to\bar{\nu}_\mu)$ for the best-fit points of cases (\textbf{A}) and (\textbf{B}) for core-crossing trajectory, respectively by blue solid and orange dashed curves.    

\begin{table}[t]
    \centering
    \caption{Best-fit values of the parameters from two fits of the short baseline data in $(3+2)$ scenario. In the last column we report the $\Delta\chi^2$ value of these best-fit points that we obtained in the analysis of IceCube data.} 
    
    \begin{tabular}{|c|c|c|c|c|c|}
    \hline 
     & $\sin^22\theta_{24}$ & $\sin^22\theta_{25}$ & $\Delta m_{41}^2$ [eV$^2$] & $\Delta m_{51}^2$ [eV$^2$] & $\Delta\chi^2$\\ \hline
    Case (\textbf{A}) from Ref.~\cite{Diaz:2019fwt} & $2.8\times10^{-2}$ &  $1.5\times10^{-2}$ & $1.32$ & $13.9$ & $4.32$ \\ \hline
     Case (\textbf{B}) from Ref.~\cite{Cianci:2017okw} & $9.1\times10^{-2}$ & $6.8\times10^{-2}$ & $0.46$ & $0.77$ & $36.43$ \\ \hline
    \end{tabular}
    \label{tab:fits}
\end{table}

Whether the best-fit points in table~\ref{tab:fits}, and the corresponding allowed regions around them, are compatible by the IceCube data can be checked by the analysis described in section~\ref{sec:analysis}. The allowed region in the 4-dimensional parameter space of $(3+2)$ scenario can be derived from the $\chi^2$ table we provide attached to this manuscript. However, to visualize this region, marginalization over two parameters out of the $(\Delta m_{41}^2,\Delta m_{51}^2,\sin^22\theta_{24},\sin^22\theta_{25})$ should be performed. Marginalization over $(\Delta m_{41}^2,\sin^22\theta_{24})$ or $(\Delta m_{51}^2,\sin^22\theta_{25})$ reduces the $(3+2)$ scenario to $(3+1)$ and in fact, as a sanity check, we have reproduced the exclusion plot of Ref.~\cite{IceCube:2016rnb} in this way. For the bounds in $(\Delta m_{41}^2,\Delta m_{51}^2)$ and $(\sin^22\theta_{24},\sin^22\theta_{25})$ planes, some caution on the marginalization procedure should be exercised. The necessary condition to obtain constraints on the mass-squared differences is $\theta_{24}\neq0$ and $\theta_{25}\neq0$, and vice versa; {\it i.e.}, to constrain $\theta_{24}$ and $\theta_{25}$, non-vanishing mass-squared differences are required. In fact, in the latter case, the mass-squared differences should lie within the $\sim (0.1,10)~{\rm eV}^2$, otherwise the IceCube data cannot provide any constraint. Thus, marginalizing over the range $0\leq\sin^22\theta_{24},\sin^22\theta_{25}\leq1$, leads to no bounds in the $(\Delta m_{41}^2,\Delta m_{51}^2)$ plane since the minimization always chooses $\sin^22\theta_{24}=\sin^22\theta_{25}=0$ which is the $3\nu$ framework. To remediate, we marginalize over the range $A\leq\sin^22\theta_{24},\sin^22\theta_{25}\leq1$, where $0\leq A\leq1$, and show the allowed region for several values of $A$\footnote{A similar issue occurs in constraining $\theta_{34}$ by the IceCube cascade events in $(3+1)$ scenario. In that case also sensitivity to $\theta_{34}$ exists when $\theta_{24}\neq0$. For a discussion of this issue and its solution, which is equivalent to our approach in this paper, see~\cite{Esmaili:2013cja}.}. 

\begin{figure}[!pt]
  \subfloat{\label{fig:dms-marg-0.02-A}
    \includegraphics[width=0.44\textwidth]{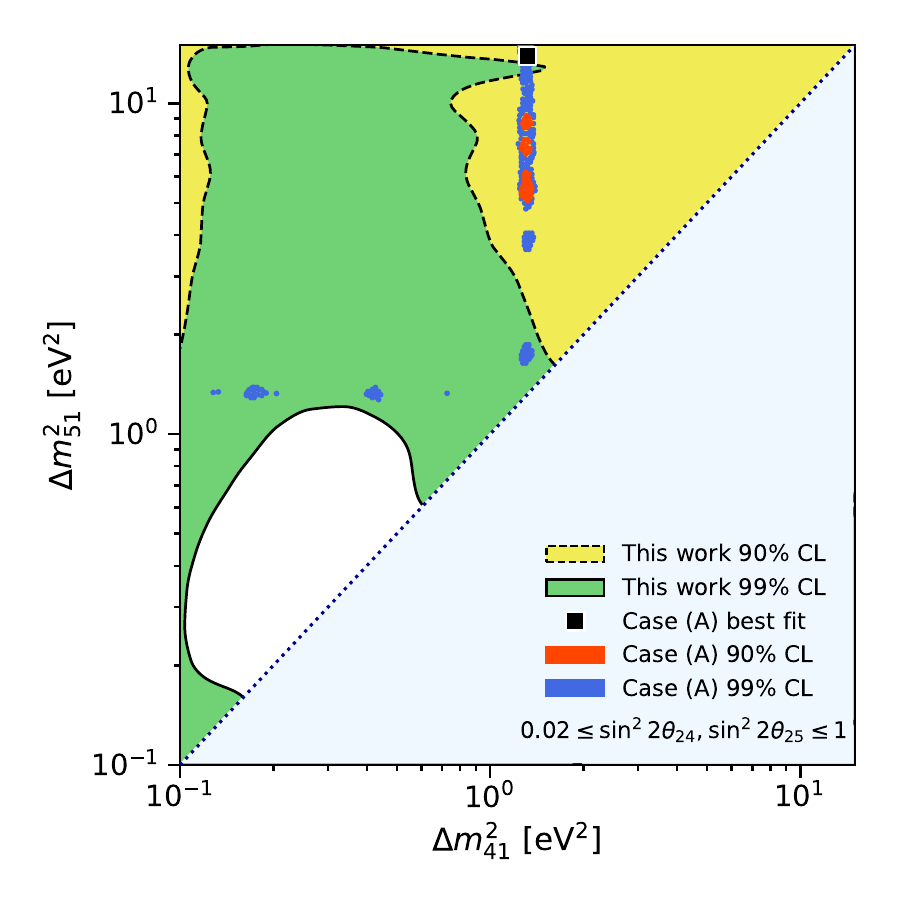}}
    \hfill
  \subfloat{\label{fig:dms-marg-0.03-A}
    \includegraphics[width=0.44\textwidth]{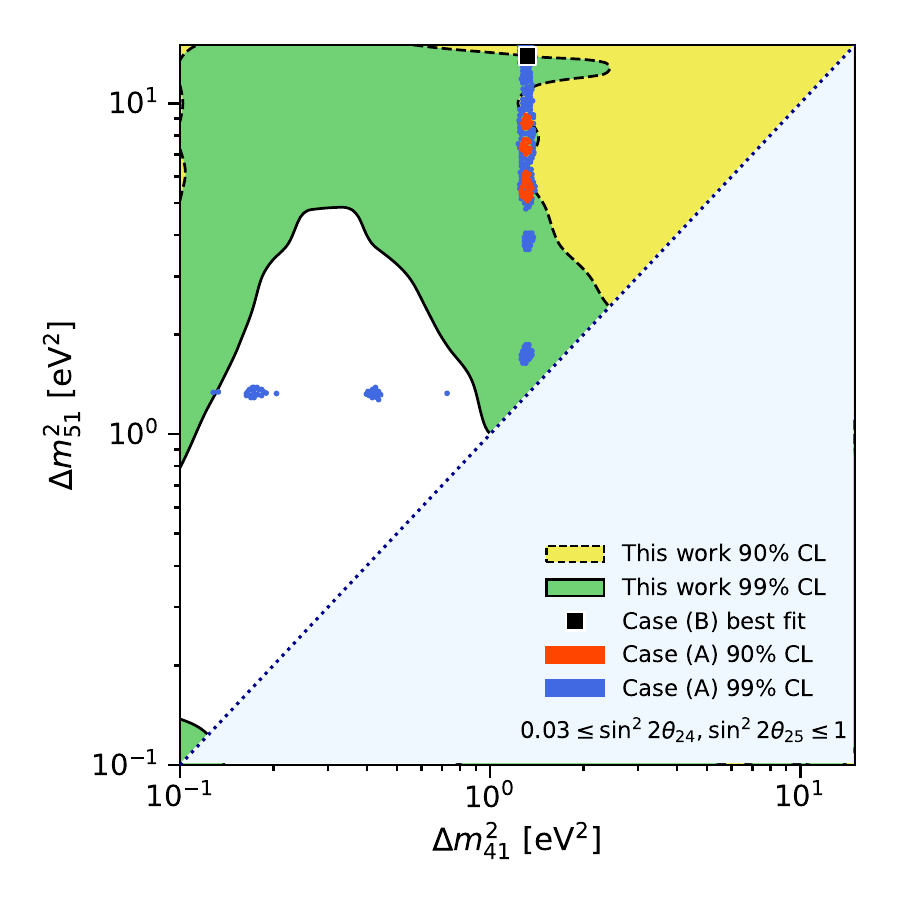}}\\
  \subfloat{\label{fig:dms-marg-0.05-A}
    \includegraphics[width=0.44\textwidth]{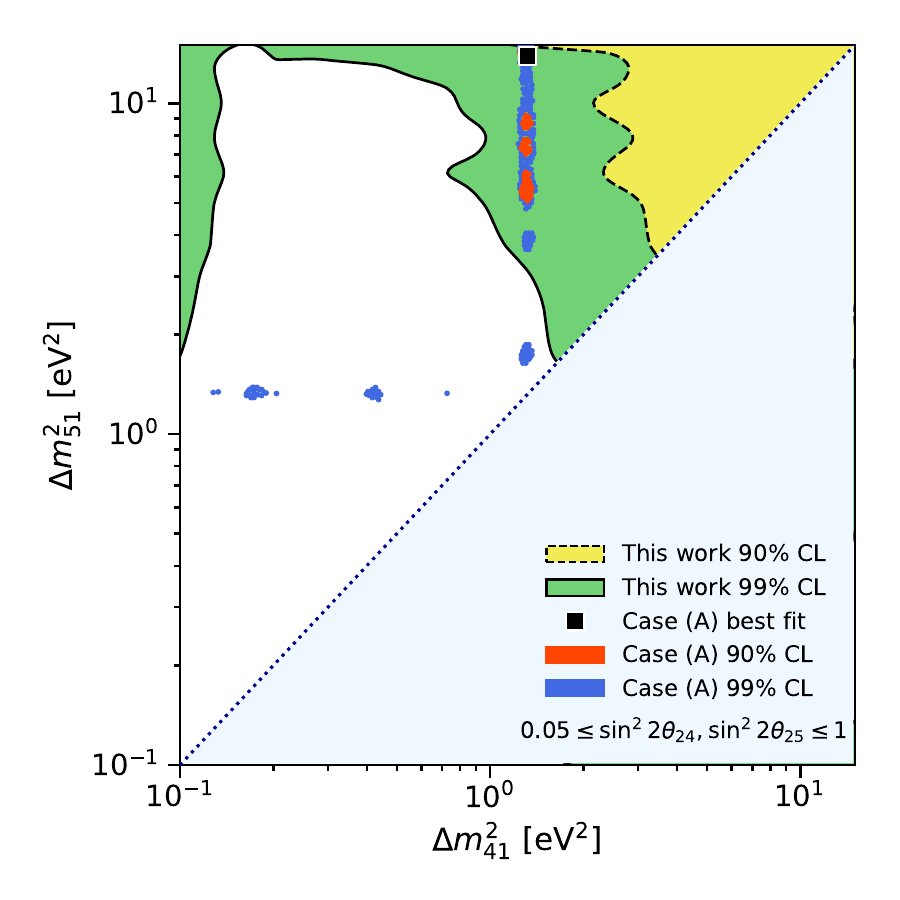}}
    \hfill
  \subfloat{\label{fig:dms-marg-0.07-A}
    \includegraphics[width=0.44\textwidth]{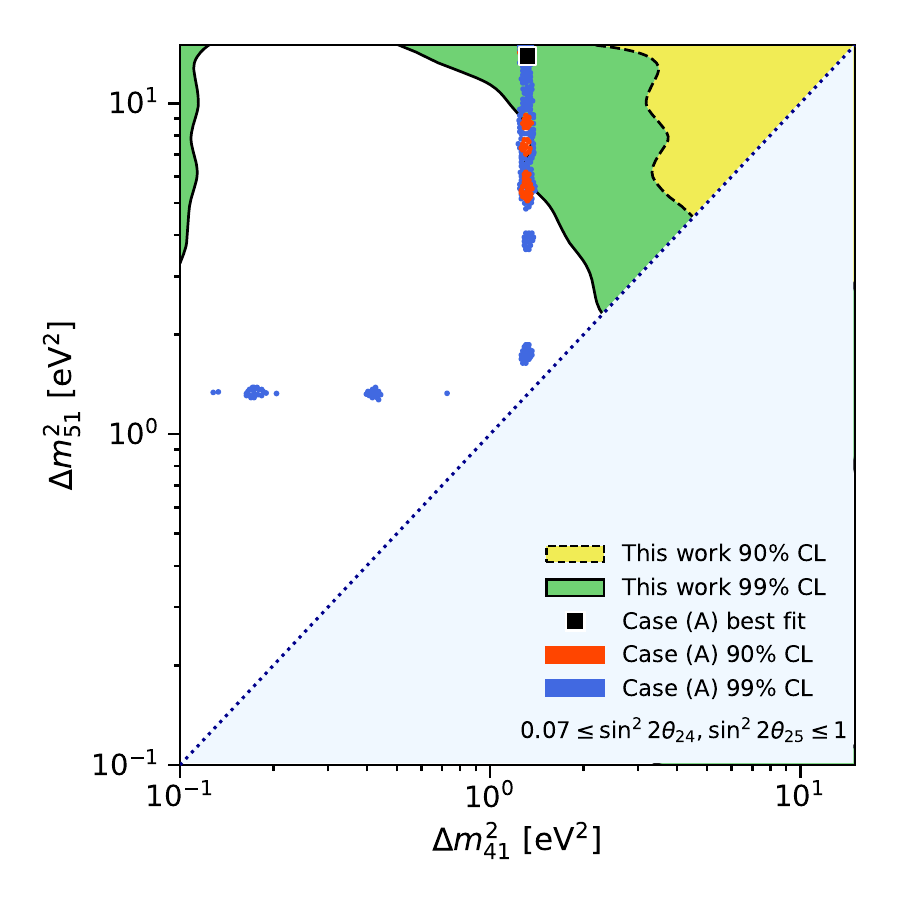}}\\
    \caption{The allowed regions by IceCube data in $(\Delta m_{41}^2,\Delta m_{51}^2)$ plane at 90\% and 99\% C.L., respectively shaded by yellow and green colors. The panels correspond to marginalization over $A\leq\sin^22\theta_{24},\sin^22\theta_{25}\leq1$ for $A=0.02$, $0.03$, $0.05$, $0.07$, $0.1$ and $0.2$. The allowed regions of case (\textbf{A}), taken from~\cite{Diaz:2019fwt}, are shown by red and blue colors, respectively at 90\% and 99\% C.L.}
    \label{fig:dms-A-allowed}
\end{figure}

\begin{figure}[!pt]
    \subfloat{\label{fig:dms-marg-0.02-B}
    \includegraphics[width=0.48\textwidth]{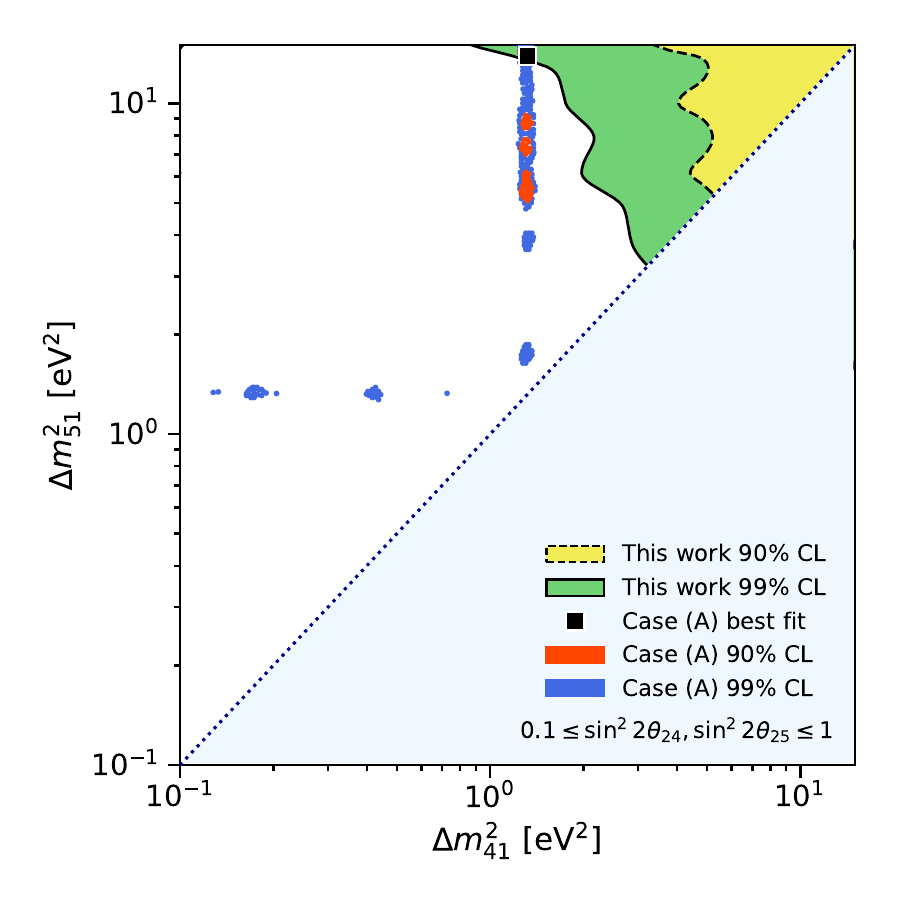}}
    \hfill
  \subfloat{\label{fig:dms-marg-0.04-B}
    \includegraphics[width=0.48\textwidth]{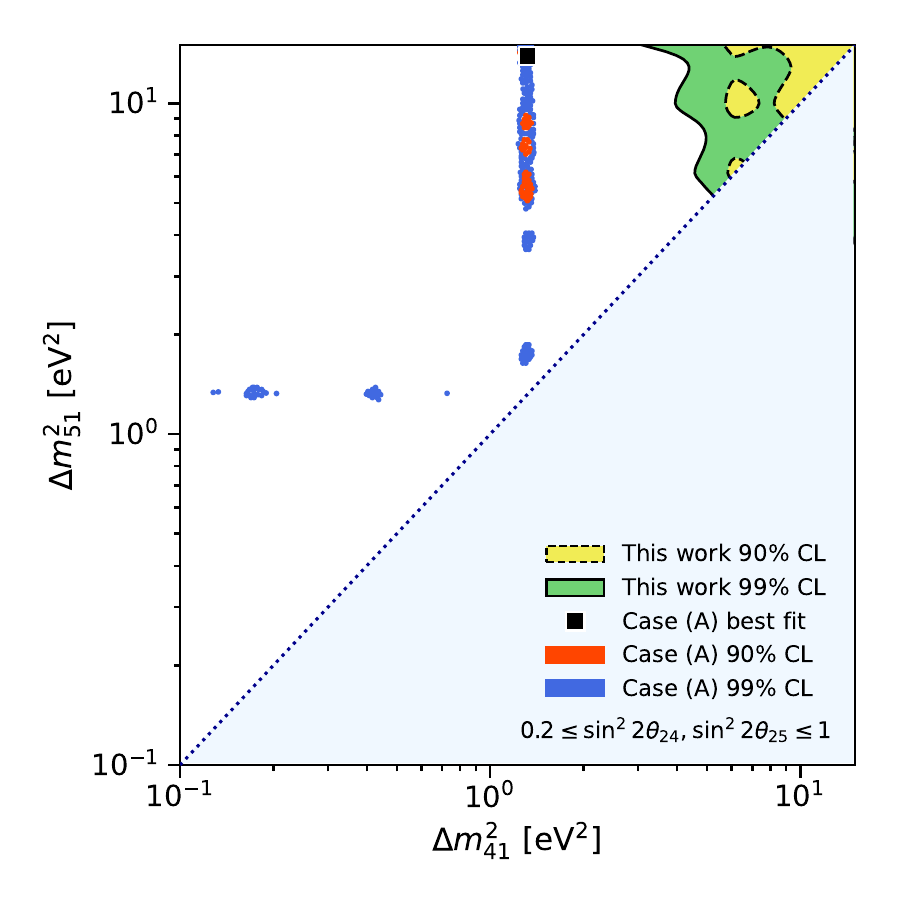}}\\
    \caption{The same as Figure~\ref{fig:dms-A-allowed} marginalizations over $0.1\leq\sin^22\theta_{24},\sin^22\theta_{25}\leq1$ in the left panel and $0.2\leq\sin^22\theta_{24},\sin^22\theta_{25}\leq1$ in the right panel.}
    \label{fig:dms-A-allowed2}
\end{figure}

\begin{figure}[h]
    \subfloat{\label{fig:dms-marg-0.02-B}
    \includegraphics[width=0.48\textwidth]{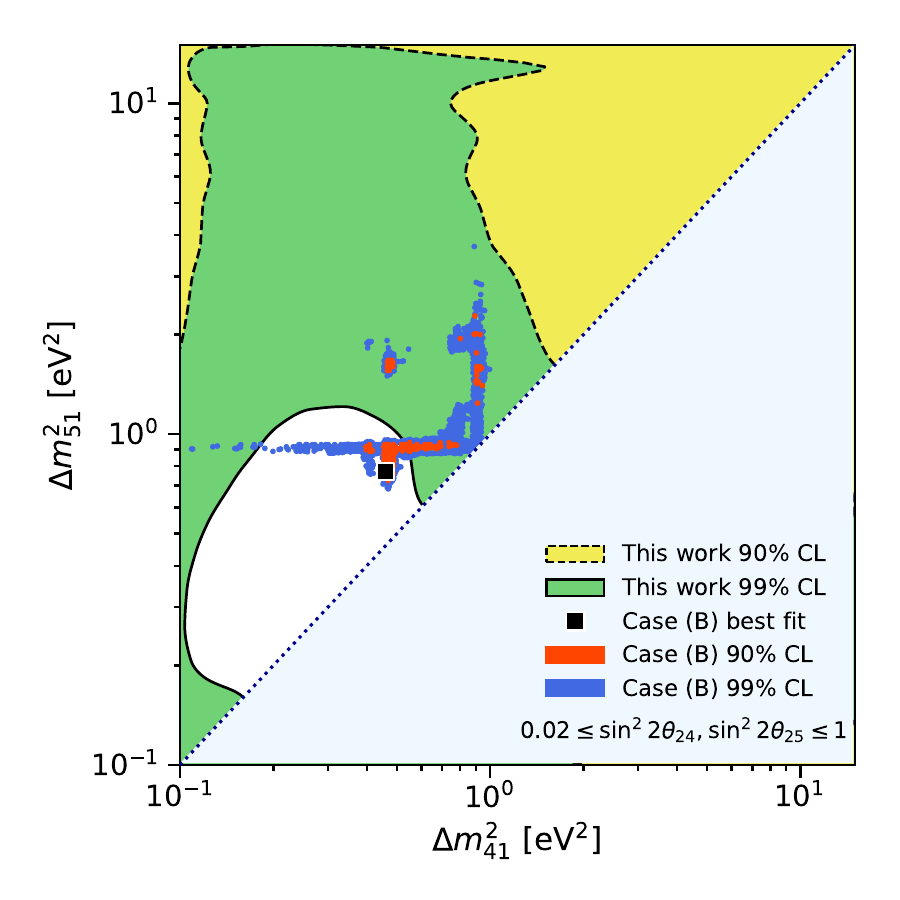}}
    \hfill
  \subfloat{\label{fig:dms-marg-0.04-B}
    \includegraphics[width=0.48\textwidth]{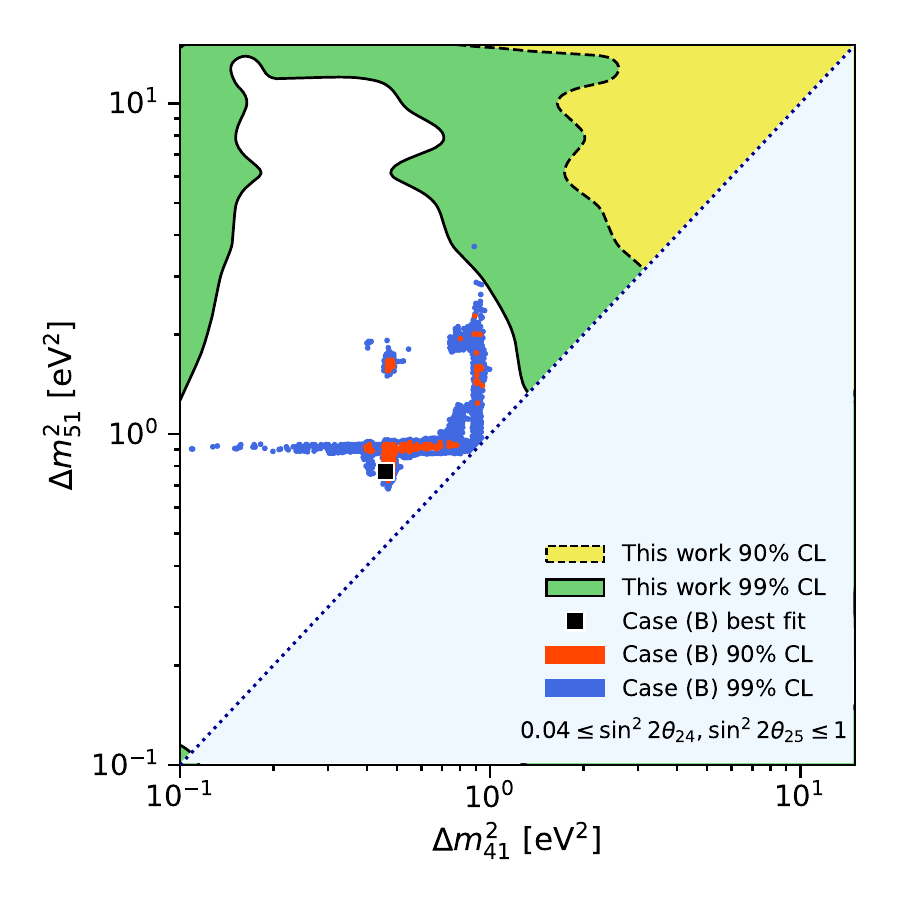}}\\
    \caption{The same as Figure~\ref{fig:dms-A-allowed} but showing the allowed regions of case (\textbf{B}), taken from~\cite{Cianci:2017okw}, and marginalizations over $0.02\leq\sin^22\theta_{24},\sin^22\theta_{25}\leq1$ in the left panel and $0.04\leq\sin^22\theta_{24},\sin^22\theta_{25}\leq1$ in the right panel.}
    \label{fig:dms-B-allowed}
\end{figure}

Figures~\ref{fig:dms-A-allowed} and \ref{fig:dms-A-allowed2} show the allowed regions by IceCube data in the $(\Delta m_{41}^2,\Delta m_{51}^2)$ plane at 90\% and 99\% C.L., respectively depicted by the shaded yellow and green regions, for $A=0.02$, $0.03$, $0.05$, $0.07$, $0.1$ and $0.2$. For comparison, we show the allowed regions of case (\textbf{A}) at 90\% and 99\% C.L. by red and blue colors, respectively. For the comparison with case (\textbf{B}), figure~\ref{fig:dms-B-allowed} shows the allowed regions for $A=0.02$ and $0.04$ compared with the allowed regions of case (\textbf{B}). From figure~\ref{fig:dms-A-allowed} we conclude that the 90\% C.L. allowed region of case (\textbf{A}) is rejected at 90\% C.L. for $\sin^22\theta_{24}$ and $\sin^22\theta_{25}\gtrsim 0.03$ by IceCube data; while the 99\% C.L. allowed region of case (\textbf{A}) at 99\% C.L. for $A=0.1$. For the case (\textbf{B}) the allowed regions at 90\% (99\%) C.L. are rejected at 90\% (99\%) C.L. for $A\simeq 0.02$ ($0.04$).

\begin{figure}[t]   
  \subfloat{\label{fig:sinq-marg-best-fitA}
    \includegraphics[width=0.44\textwidth]{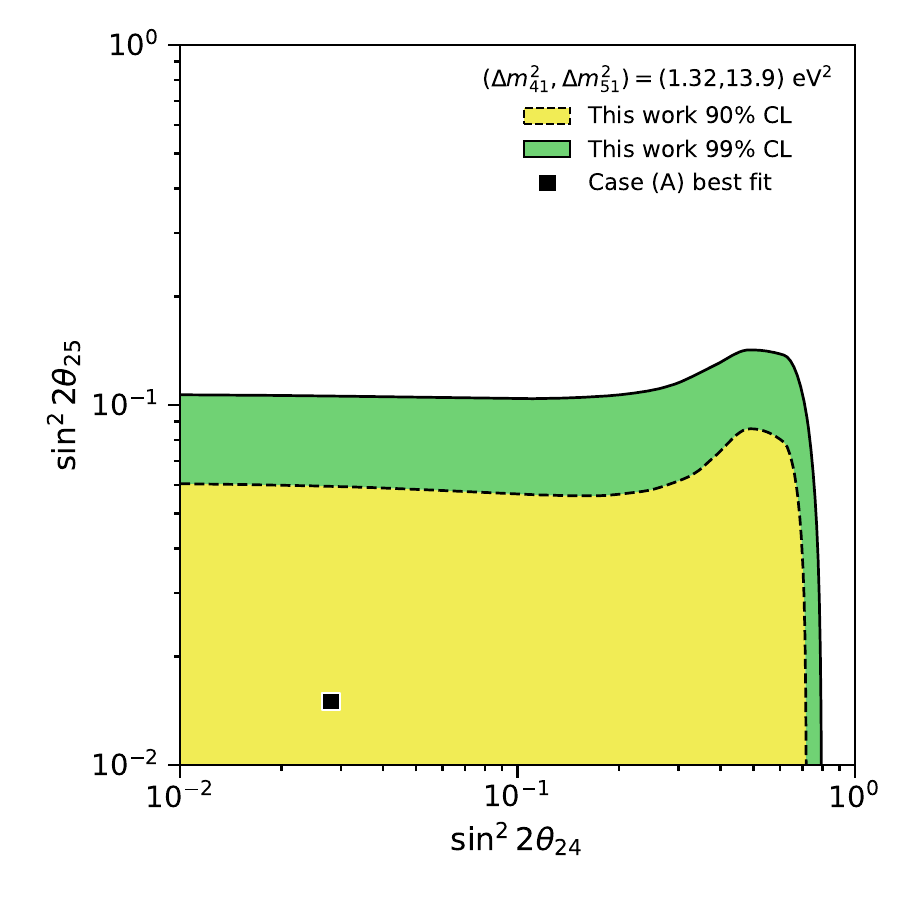}}
    \hfill
  \subfloat{\label{fig:sinq-marg-best-fitB}
    \includegraphics[width=0.44\textwidth]{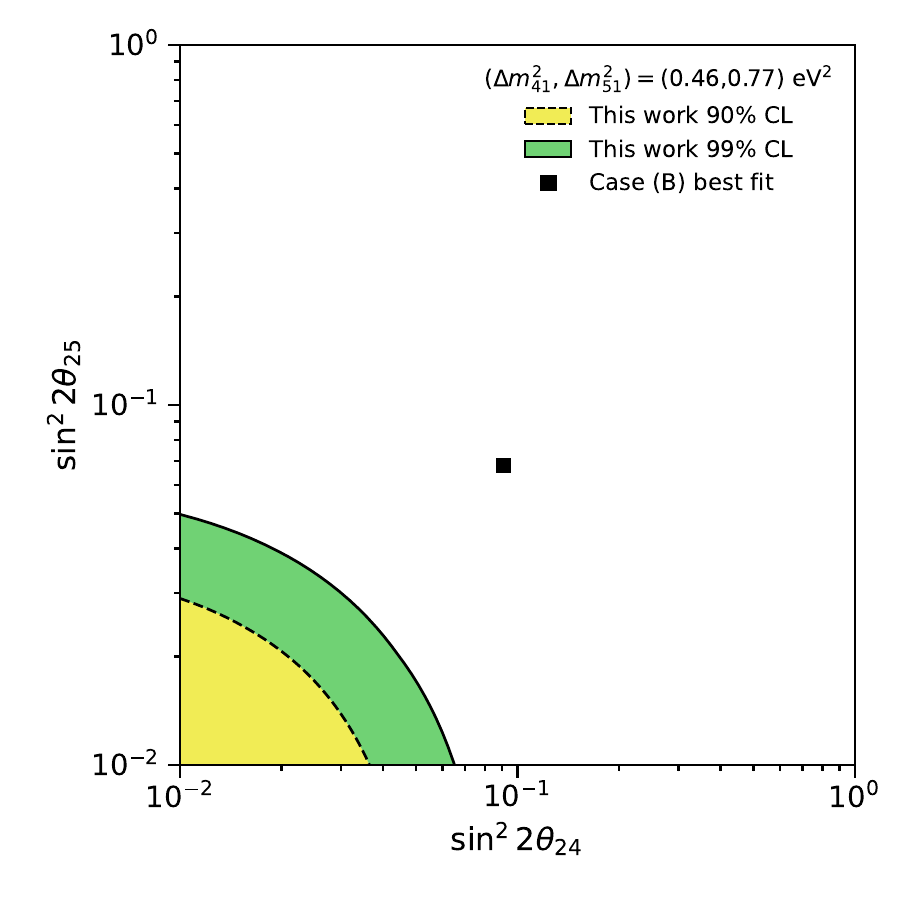}}\\
  \subfloat{\label{fig:sinq-marg-0.5-5}
    \includegraphics[width=0.44\textwidth]{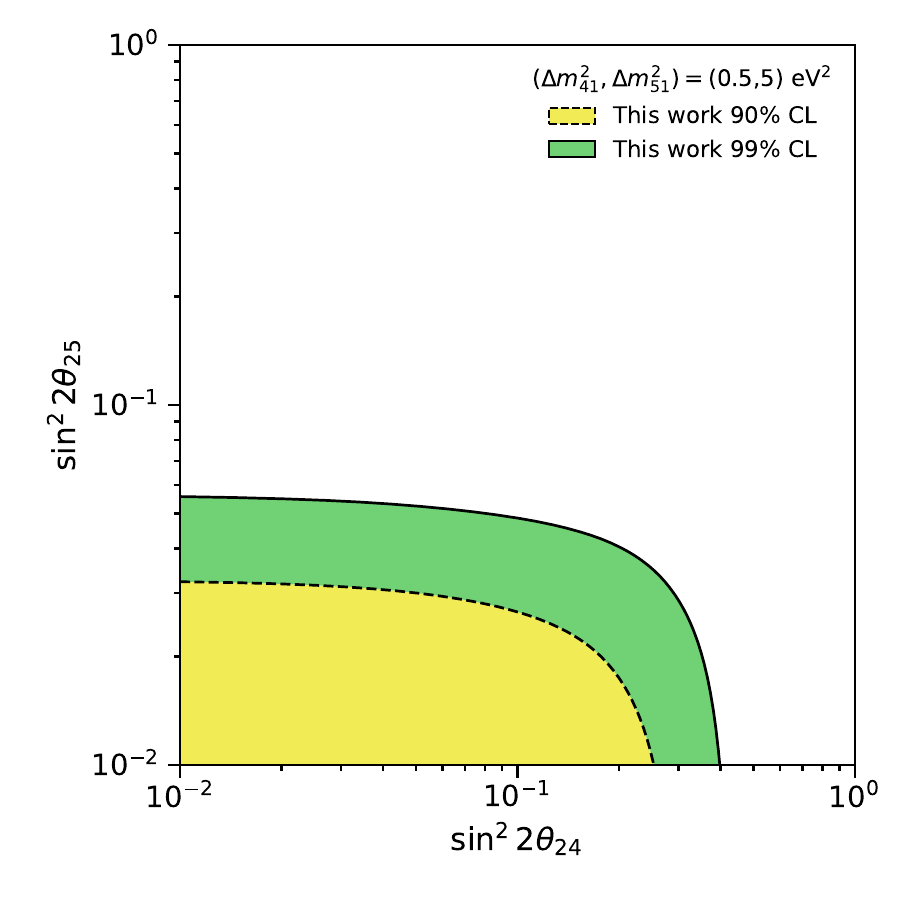}}
    \hfill
   \subfloat{\label{fig:sinq-marg-1-1}
    \includegraphics[width=0.44\textwidth]{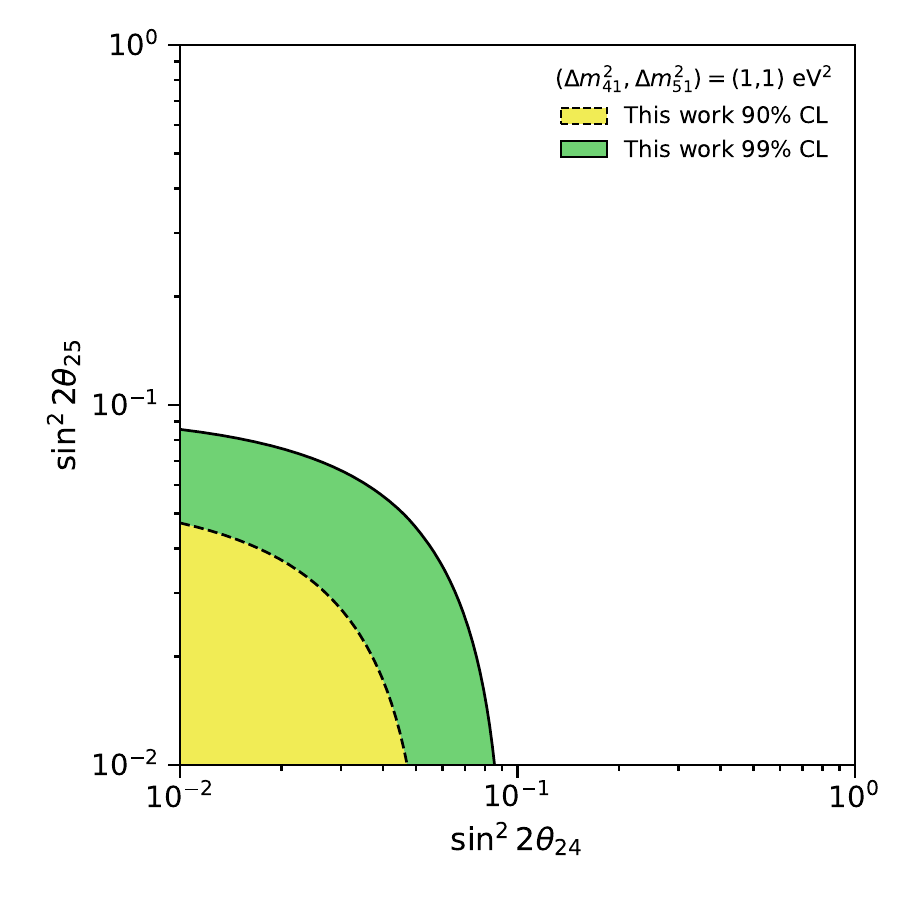}}\\
    \caption{The allowed regions, for $(3+2)$ scenario, by IceCube data in $(\sin^22\theta_{24},\sin^22\theta_{25})$ plane at 90\% and 99\% C.L., respectively shaded by yellow and green colors. In the upper left and right panels the mass-squared differences are fixed to the best-fit points of case (\textbf{A}) and (\textbf{B}), respectively (see table~\ref{tab:fits}). In the lower left and right panels we choose $(\Delta m_{41}^2,\Delta m_{51}^2)= (0.5,5)~{\rm eV}^2$ and  $(1,1)~{\rm eV}^2$, respectively.}
   \label{fig:sinq-allowed}
\end{figure}

\begin{figure}[!pt]
    \subfloat{\label{fig:sinq-marg-1-2}
    \includegraphics[width=0.48\textwidth]{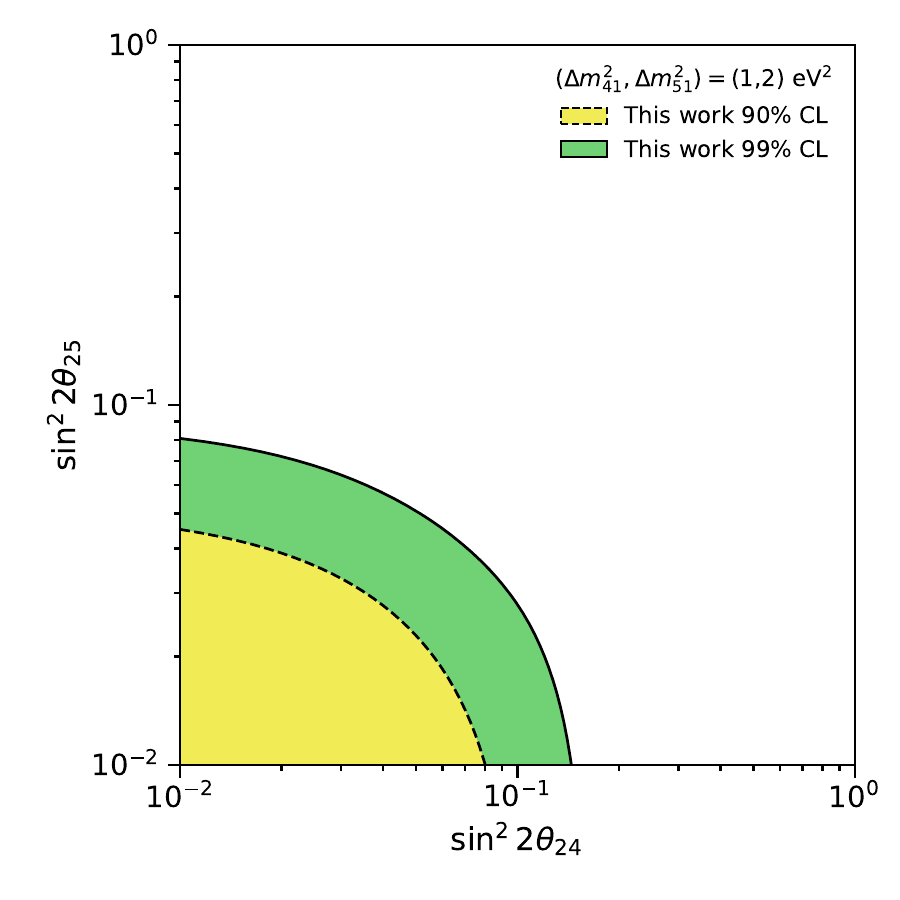}}
    \hfill
  \subfloat{\label{fig:sinq-marg-1-5}
    \includegraphics[width=0.48\textwidth]{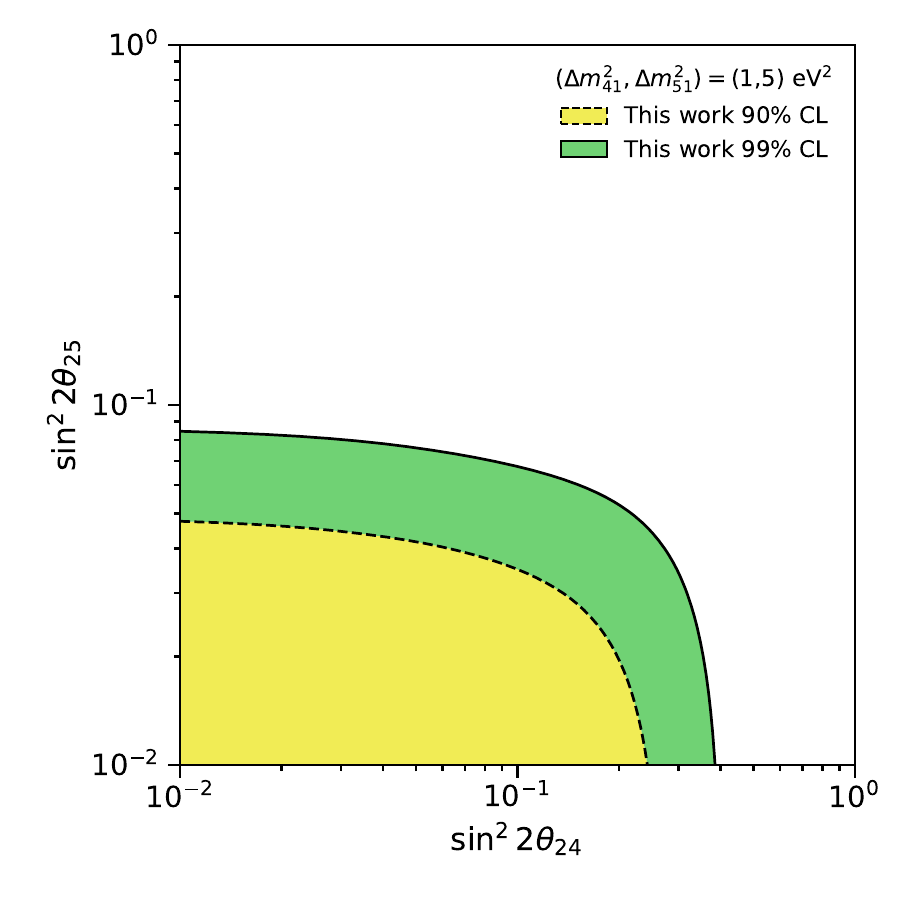}}\\
    \caption{The same as Figure~\ref{fig:sinq-allowed} for $(\Delta m_{41}^2,\Delta m_{51}^2)= (1,2)~{\rm eV}^2$, in the left panel, and $(1,5)~{\rm eV}^2$ in the right panel.}
    \label{fig:sinq-allowed2}
\end{figure}

In the $(\sin^22\theta_{24},\sin^22\theta_{25})$ plane the allowed region can be derived by marginalization over $\Delta m_{41}^2$ and $\Delta m_{51}^2$. Once more, taking a range like $[0.1,16]~{\rm eV}^2$ for the marginalization over mass-squared differences leads to no limit, since mass-squared differences $\sim0.1~{\rm eV}^2$ and $\sim16~{\rm eV}^2$ lead to resonances in oscillation probability outside of the energy range of IceCube data. In fact, for large values of mass-squared difference the resonances are at high energies (for $\Delta m^2=10~{\rm eV}^2$, the parametric resonance is at $\sim23$~TeV and the MSW at $\sim40$~TeV) where the atmospheric neutrino flux is small. In principle, by larger detectors such as IceCube-Gen2~\cite{IceCube-Gen2:2020qha} and several years of data collection, large values of mass-squared differences can be probed. For small mass-squared differences, the limitation originates from the lower energy cut on the IceCube data, $400$~GeV for the dataset of our analysis, and data release by IceCube at lower energies can lift this limitation (see~\cite{Razzaque:2012tp}). Taking into account these considerations, we derive the bounds on $(\sin^22\theta_{24}$ and $\sin^22\theta_{25})$ for fixed values of $\Delta m_{41}^2$ and $\Delta m_{51}^2$. Figures~\ref{fig:sinq-allowed} \ref{fig:sinq-allowed2} show these bounds for several choices of mass-squared differences, including the best-fit values of cases (\textbf{A}) and (\textbf{B}), respectively in the upper left and right panels of figure~\ref{fig:sinq-allowed}. The shaded yellow and green regions depict the allowed regions at 90\% and 99\% C.L., respectively. The rejection of the best-fit point of case (\textbf{B}) is evident from the upper right panel of figure~\ref{fig:sinq-allowed}, while the upper left panel shows the compatibility of the best-fit point of case (\textbf{A}) with the IceCube limits.

\section{Summary and Conclusions\label{sec:SumConc}}

Several anomalies observed in the short baseline, reactor and solar experiments are hinting at the existence of one or more sterile neutrino states. While the mass scale of these states should be $\sim1$~eV for the interpretation of anomalies, the number of these states is practically free and can be one, two, or three for aesthetic reason of matching with the three active neutrino states. In fact, larger number of sterile neutrino states, due to extra interference terms, generally results in better fit to the features of the observed spectra in anomalous data sets, though at the price of dealing with a considerably larger number of parameters in the fit. Even though these improvements of the fits motivate the consideration of larger number of sterile neutrino states, there are experiments that the effect of additional sterile states is the contrary; \textit{i.e.}, more sterile states leads to stronger bounds on the parameter space. An example is the effect of the sterile neutrinos on the high energy atmospheric neutrinos observable at IceCube experiment. Since the sterile neutrino introduces resonances in the $\bar{\nu}_\mu\to\bar{\nu}_\mu$ oscillation probability, existence of more sterile states means more resonances and hence stronger bounds. 

We studied the $(3+2)$ scenario which contains two additional sterile neutrinos. We argued that out of the 20 parameters in this scenario, only 4 parameters ($\Delta m_{41}^2$, $\Delta m_{51}^2$, $\sin^22\theta_{24}$, $\sin^22\theta_{25}$) need consideration if one is interested in the most conservative bound on this scenario. By analyzing the one-year atmospheric neutrino data set of IceCube collected during the 2011-2012, bounds on the 4-dimensional parameter space and their projections on the ($\Delta m_{41}^2$, $\Delta m_{51}^2$) and ($\sin^22\theta_{24}$, $\sin^22\theta_{25}$) planes have been derived. We performed a binned Poisson likelihood analysis and took into account various systematic and statistical uncertainties, including the uncertainties on the detector efficiency and variants of atmospheric neutrino flux. To compare the obtained bounds with the allowed regions from short baseline experiments, we chose two different fits from available analyses in the literature, named cases (\textbf{A}) and (\textbf{B}). The obtained bounds show that while the allowed regions of case (\textbf{A}) seem to be compatible with the IceCube data, the allowed regions of case (\textbf{B}) are in strong tension with the derived bounds. However, to derive the precise level of compatibility or rejection of these allowed regions, and in fact to make a meaningful fit, a joint analysis of IceCube and short baseline data should be performed. To this end, we provide a table of $\chi^2$ values for a grid of points in the parameter space. 

Finally, we would mention that other data sets with higher statistics have been published by IceCube collaboration, such as~\cite{IceCube:2020phf} which contains eight years of data and \cite{IceCube:2024kel} including 10.7 years of data. Comparing the result of~\cite{IceCube:2020phf} with \cite{IceCube:2016rnb} which we used, shows that the exclusion curves are close. Ref.~\cite{IceCube:2024kel} shows significant improvement with respect to \cite{IceCube:2016rnb}. We opted for the one-year data set since more information on the uncertainties have been provided for this data set which makes a comprehensive analysis feasible, while for the other two data sets these information are not available.

\begin{acknowledgments}
E.~C. acknowledges support by the Conselho Nacional de Desenvolvimento Cient\'{i}fico e Tecnol\'{o}gico (CNPq) scholarship No. 140121/2022-6. A.~E. and A.~A.~Q. thank partial financial support by the Brazilian funding agency CNPq (grant 407149/2021). The authors thank John David Rogers Computing Center (CCJDR) in the Institute of Physics “Gleb Wataghin”, University of Campinas, for providing computing resources.
\end{acknowledgments}

\bibliographystyle{JHEP}
\bibliography{refs}

\hyphenation{Post-Script Sprin-ger}

\providecommand{\href}[2]{#2}\begingroup\raggedright\begin{thebibliography}{10}

\bibitem{Hyper-Kamiokande:2018ofw}
{\scshape Hyper-Kamiokande} collaboration, \emph{{Hyper-Kamiokande Design
  Report}},  \href{https://arxiv.org/abs/1805.04163}{{\ttfamily 1805.04163}}.

\bibitem{DUNE:2020ypp}
{\scshape DUNE} collaboration, \emph{{Deep Underground Neutrino Experiment
  (DUNE), Far Detector Technical Design Report, Volume II: DUNE Physics}},
  \href{https://arxiv.org/abs/2002.03005}{{\ttfamily 2002.03005}}.

\bibitem{JUNO:2015zny}
{\scshape JUNO} collaboration, \emph{{Neutrino Physics with JUNO}},
  \href{https://doi.org/10.1088/0954-3899/43/3/030401}{\emph{J. Phys. G}
  {\bfseries 43} (2016) 030401}
  [\href{https://arxiv.org/abs/1507.05613}{{\ttfamily 1507.05613}}].

\bibitem{Esteban:2020cvm}
I.~Esteban, M.C.~Gonzalez-Garcia, M.~Maltoni, T.~Schwetz and A.~Zhou,
  \emph{{The fate of hints: updated global analysis of three-flavor neutrino
  oscillations}}, \href{https://doi.org/10.1007/JHEP09(2020)178}{\emph{JHEP}
  {\bfseries 09} (2020) 178}
  [\href{https://arxiv.org/abs/2007.14792}{{\ttfamily 2007.14792}}].

\bibitem{deSalas:2020pgw}
P.F.~de~Salas, D.V.~Forero, S.~Gariazzo, P.~Mart\'\i{}nez-Mirav\'e, O.~Mena,
  C.A.~Ternes et~al., \emph{{2020 global reassessment of the neutrino
  oscillation picture}},
  \href{https://doi.org/10.1007/JHEP02(2021)071}{\emph{JHEP} {\bfseries 02}
  (2021) 071} [\href{https://arxiv.org/abs/2006.11237}{{\ttfamily
  2006.11237}}].

\bibitem{Capozzi:2018ubv}
F.~Capozzi, E.~Lisi, A.~Marrone and A.~Palazzo, \emph{{Current unknowns in the
  three neutrino framework}},
  \href{https://doi.org/10.1016/j.ppnp.2018.05.005}{\emph{Prog. Part. Nucl.
  Phys.} {\bfseries 102} (2018) 48}
  [\href{https://arxiv.org/abs/1804.09678}{{\ttfamily 1804.09678}}].

\bibitem{LSND:2001aii}
{\scshape LSND} collaboration, \emph{{Evidence for neutrino oscillations from
  the observation of $\bar{\nu}_e$ appearance in a $\bar{\nu}_\mu$ beam}},
  \href{https://doi.org/10.1103/PhysRevD.64.112007}{\emph{Phys. Rev. D}
  {\bfseries 64} (2001) 112007}
  [\href{https://arxiv.org/abs/hep-ex/0104049}{{\ttfamily hep-ex/0104049}}].

\bibitem{MiniBooNE:2020pnu}
{\scshape MiniBooNE} collaboration, \emph{{Updated MiniBooNE neutrino
  oscillation results with increased data and new background studies}},
  \href{https://doi.org/10.1103/PhysRevD.103.052002}{\emph{Phys. Rev. D}
  {\bfseries 103} (2021) 052002}
  [\href{https://arxiv.org/abs/2006.16883}{{\ttfamily 2006.16883}}].

\bibitem{Mueller:2011nm}
T.A.~Mueller et~al., \emph{{Improved Predictions of Reactor Antineutrino
  Spectra}}, \href{https://doi.org/10.1103/PhysRevC.83.054615}{\emph{Phys. Rev.
  C} {\bfseries 83} (2011) 054615}
  [\href{https://arxiv.org/abs/1101.2663}{{\ttfamily 1101.2663}}].

\bibitem{Huber:2011wv}
P.~Huber, \emph{{On the determination of anti-neutrino spectra from nuclear
  reactors}}, \href{https://doi.org/10.1103/PhysRevC.85.029901}{\emph{Phys.
  Rev. C} {\bfseries 84} (2011) 024617}
  [\href{https://arxiv.org/abs/1106.0687}{{\ttfamily 1106.0687}}].

\bibitem{DANSS:2018fnn}
{\scshape DANSS} collaboration, \emph{{Search for sterile neutrinos at the
  DANSS experiment}},
  \href{https://doi.org/10.1016/j.physletb.2018.10.038}{\emph{Phys. Lett. B}
  {\bfseries 787} (2018) 56}
  [\href{https://arxiv.org/abs/1804.04046}{{\ttfamily 1804.04046}}].

\bibitem{NEOS:2016wee}
{\scshape NEOS} collaboration, \emph{{Sterile Neutrino Search at the NEOS
  Experiment}},
  \href{https://doi.org/10.1103/PhysRevLett.118.121802}{\emph{Phys. Rev. Lett.}
  {\bfseries 118} (2017) 121802}
  [\href{https://arxiv.org/abs/1610.05134}{{\ttfamily 1610.05134}}].

\bibitem{Serebrov:2020kmd}
A.P.~Serebrov et~al., \emph{{Search for sterile neutrinos with the Neutrino-4
  experiment and measurement results}},
  \href{https://doi.org/10.1103/PhysRevD.104.032003}{\emph{Phys. Rev. D}
  {\bfseries 104} (2021) 032003}
  [\href{https://arxiv.org/abs/2005.05301}{{\ttfamily 2005.05301}}].

\bibitem{PROSPECT:2020sxr}
{\scshape PROSPECT} collaboration, \emph{{Improved short-baseline neutrino
  oscillation search and energy spectrum measurement with the PROSPECT
  experiment at HFIR}},
  \href{https://doi.org/10.1103/PhysRevD.103.032001}{\emph{Phys. Rev. D}
  {\bfseries 103} (2021) 032001}
  [\href{https://arxiv.org/abs/2006.11210}{{\ttfamily 2006.11210}}].

\bibitem{STEREO:2019ztb}
{\scshape STEREO} collaboration, \emph{{Improved sterile neutrino constraints
  from the STEREO experiment with 179 days of reactor-on data}},
  \href{https://doi.org/10.1103/PhysRevD.102.052002}{\emph{Phys. Rev. D}
  {\bfseries 102} (2020) 052002}
  [\href{https://arxiv.org/abs/1912.06582}{{\ttfamily 1912.06582}}].

\bibitem{Giunti:2021iti}
C.~Giunti, Y.F.~Li, C.A.~Ternes and Y.Y.~Zhang, \emph{{Neutrino-4 anomaly:
  oscillations or fluctuations?}},
  \href{https://doi.org/10.1016/j.physletb.2021.136214}{\emph{Phys. Lett. B}
  {\bfseries 816} (2021) 136214}
  [\href{https://arxiv.org/abs/2101.06785}{{\ttfamily 2101.06785}}].

\bibitem{SAGE:2009eeu}
{\scshape SAGE} collaboration, \emph{{Measurement of the solar neutrino capture
  rate with gallium metal. III: Results for the 2002--2007 data-taking
  period}}, \href{https://doi.org/10.1103/PhysRevC.80.015807}{\emph{Phys. Rev.
  C} {\bfseries 80} (2009) 015807}
  [\href{https://arxiv.org/abs/0901.2200}{{\ttfamily 0901.2200}}].

\bibitem{Kaether:2010ag}
F.~Kaether, W.~Hampel, G.~Heusser, J.~Kiko and T.~Kirsten, \emph{{Reanalysis of
  the GALLEX solar neutrino flux and source experiments}},
  \href{https://doi.org/10.1016/j.physletb.2010.01.030}{\emph{Phys. Lett. B}
  {\bfseries 685} (2010) 47} [\href{https://arxiv.org/abs/1001.2731}{{\ttfamily
  1001.2731}}].

\bibitem{Barinov:2021asz}
V.V.~Barinov et~al., \emph{{Results from the Baksan Experiment on Sterile
  Transitions (BEST)}},
  \href{https://doi.org/10.1103/PhysRevLett.128.232501}{\emph{Phys. Rev. Lett.}
  {\bfseries 128} (2022) 232501}
  [\href{https://arxiv.org/abs/2109.11482}{{\ttfamily 2109.11482}}].

\bibitem{Elliott:2023cvh}
S.R.~Elliott, V.~Gavrin and W.~Haxton, \emph{{The gallium anomaly}},
  \href{https://doi.org/10.1016/j.ppnp.2023.104082}{\emph{Prog. Part. Nucl.
  Phys.} {\bfseries 134} (2024) 104082}
  [\href{https://arxiv.org/abs/2306.03299}{{\ttfamily 2306.03299}}].

\bibitem{Giunti:2022btk}
C.~Giunti, Y.F.~Li, C.A.~Ternes, O.~Tyagi and Z.~Xin, \emph{{Gallium Anomaly:
  critical view from the global picture of \ensuremath{\nu}$_{e}$ and $
  {\overline{\nu}}_e $ disappearance}},
  \href{https://doi.org/10.1007/JHEP10(2022)164}{\emph{JHEP} {\bfseries 10}
  (2022) 164} [\href{https://arxiv.org/abs/2209.00916}{{\ttfamily
  2209.00916}}].

\bibitem{Liu:1997yb}
Q.Y.~Liu and A.Y.~Smirnov, \emph{{Neutrino mass spectrum with muon-neutrino
  ---\ensuremath{>} sterile-neutrino oscillations of atmospheric neutrinos}},
  \href{https://doi.org/10.1016/S0550-3213(98)00269-7}{\emph{Nucl. Phys. B}
  {\bfseries 524} (1998) 505}
  [\href{https://arxiv.org/abs/hep-ph/9712493}{{\ttfamily hep-ph/9712493}}].

\bibitem{Liu:1998nb}
Q.Y.~Liu, S.P.~Mikheyev and A.Y.~Smirnov, \emph{{Parametric resonance in
  oscillations of atmospheric neutrinos?}},
  \href{https://doi.org/10.1016/S0370-2693(98)01102-2}{\emph{Phys. Lett. B}
  {\bfseries 440} (1998) 319}
  [\href{https://arxiv.org/abs/hep-ph/9803415}{{\ttfamily hep-ph/9803415}}].

\bibitem{Nunokawa:2003ep}
H.~Nunokawa, O.L.G.~Peres and R.~Zukanovich~Funchal, \emph{{Probing the LSND
  mass scale and four neutrino scenarios with a neutrino telescope}},
  \href{https://doi.org/10.1016/S0370-2693(03)00603-8}{\emph{Phys. Lett. B}
  {\bfseries 562} (2003) 279}
  [\href{https://arxiv.org/abs/hep-ph/0302039}{{\ttfamily hep-ph/0302039}}].

\bibitem{Yasuda:2000xs}
O.~Yasuda, \emph{{Neutrino oscillations with four generations}},  in
  \emph{{Joint U.S. / Japan Workshop on New Initiatives in Muon Lepton Flavor
  Violation and Neutrino Oscillation with High Intense Muon and Neutrino
  Sources}}, pp.~151--163, 10, 2000,
  \href{https://doi.org/10.1142/9789812777003_0017}{DOI}
  [\href{https://arxiv.org/abs/hep-ph/0102166}{{\ttfamily hep-ph/0102166}}].

\bibitem{Choubey:2007ji}
S.~Choubey, \emph{{Signature of sterile species in atmospheric neutrino data at
  neutrino telescopes}},
  \href{https://doi.org/10.1088/1126-6708/2007/12/014}{\emph{JHEP} {\bfseries
  12} (2007) 014} [\href{https://arxiv.org/abs/0709.1937}{{\ttfamily
  0709.1937}}].

\bibitem{Barger:2011rc}
V.~Barger, Y.~Gao and D.~Marfatia, \emph{{Is there evidence for sterile
  neutrinos in IceCube data?}},
  \href{https://doi.org/10.1103/PhysRevD.85.011302}{\emph{Phys. Rev. D}
  {\bfseries 85} (2012) 011302}
  [\href{https://arxiv.org/abs/1109.5748}{{\ttfamily 1109.5748}}].

\bibitem{Razzaque:2011ab}
S.~Razzaque and A.Y.~Smirnov, \emph{{Searching for sterile neutrinos in ice}},
  \href{https://doi.org/10.1007/JHEP07(2011)084}{\emph{JHEP} {\bfseries 07}
  (2011) 084} [\href{https://arxiv.org/abs/1104.1390}{{\ttfamily 1104.1390}}].

\bibitem{Razzaque:2012tp}
S.~Razzaque and A.Y.~Smirnov, \emph{{Searches for sterile neutrinos with
  IceCube DeepCore}},
  \href{https://doi.org/10.1103/PhysRevD.85.093010}{\emph{Phys. Rev. D}
  {\bfseries 85} (2012) 093010}
  [\href{https://arxiv.org/abs/1203.5406}{{\ttfamily 1203.5406}}].

\bibitem{Esmaili:2012nz}
A.~Esmaili, F.~Halzen and O.L.G.~Peres, \emph{{Constraining Sterile Neutrinos
  with AMANDA and IceCube Atmospheric Neutrino Data}},
  \href{https://doi.org/10.1088/1475-7516/2012/11/041}{\emph{JCAP} {\bfseries
  11} (2012) 041} [\href{https://arxiv.org/abs/1206.6903}{{\ttfamily
  1206.6903}}].

\bibitem{Esmaili:2013vza}
A.~Esmaili and A.Y.~Smirnov, \emph{{Restricting the LSND and MiniBooNE sterile
  neutrinos with the IceCube atmospheric neutrino data}},
  \href{https://doi.org/10.1007/JHEP12(2013)014}{\emph{JHEP} {\bfseries 12}
  (2013) 014} [\href{https://arxiv.org/abs/1307.6824}{{\ttfamily 1307.6824}}].

\bibitem{Esmaili:2013cja}
A.~Esmaili, F.~Halzen and O.L.G.~Peres, \emph{{Exploring $\nu_\tau - \nu_s$
  mixing with cascade events in DeepCore}},
  \href{https://doi.org/10.1088/1475-7516/2013/07/048}{\emph{JCAP} {\bfseries
  07} (2013) 048} [\href{https://arxiv.org/abs/1303.3294}{{\ttfamily
  1303.3294}}].

\bibitem{IceCube:2016rnb}
{\scshape IceCube} collaboration, \emph{{Searches for Sterile Neutrinos with
  the IceCube Detector}},
  \href{https://doi.org/10.1103/PhysRevLett.117.071801}{\emph{Phys. Rev. Lett.}
  {\bfseries 117} (2016) 071801}
  [\href{https://arxiv.org/abs/1605.01990}{{\ttfamily 1605.01990}}].

\bibitem{IceCube:2020phf}
{\scshape IceCube} collaboration, \emph{{eV-Scale Sterile Neutrino Search Using
  Eight Years of Atmospheric Muon Neutrino Data from the IceCube Neutrino
  Observatory}},
  \href{https://doi.org/10.1103/PhysRevLett.125.141801}{\emph{Phys. Rev. Lett.}
  {\bfseries 125} (2020) 141801}
  [\href{https://arxiv.org/abs/2005.12942}{{\ttfamily 2005.12942}}].

\bibitem{Diaz:2019fwt}
A.~Diaz, C.A.~Arg\"uelles, G.H.~Collin, J.M.~Conrad and M.H.~Shaevitz,
  \emph{{Where Are We With Light Sterile Neutrinos?}},
  \href{https://doi.org/10.1016/j.physrep.2020.08.005}{\emph{Phys. Rept.}
  {\bfseries 884} (2020) 1} [\href{https://arxiv.org/abs/1906.00045}{{\ttfamily
  1906.00045}}].

\bibitem{Cianci:2017okw}
D.~Cianci, A.~Furmanski, G.~Karagiorgi and M.~Ross-Lonergan, \emph{{Prospects
  of Light Sterile Neutrino Oscillation and CP Violation Searches at the
  Fermilab Short Baseline Neutrino Facility}},
  \href{https://doi.org/10.1103/PhysRevD.96.055001}{\emph{Phys. Rev. D}
  {\bfseries 96} (2017) 055001}
  [\href{https://arxiv.org/abs/1702.01758}{{\ttfamily 1702.01758}}].

\bibitem{Hardin:2022muu}
J.M.~Hardin, I.~Martinez-Soler, A.~Diaz, M.~Jin, N.W.~Kamp, C.A.~Arg\"uelles
  et~al., \emph{{New Clues about light sterile neutrinos: preference for models
  with damping effects in global fits}},
  \href{https://doi.org/10.1007/JHEP09(2023)058}{\emph{JHEP} {\bfseries 09}
  (2023) 058} [\href{https://arxiv.org/abs/2211.02610}{{\ttfamily
  2211.02610}}].

\bibitem{Acero:2022wqg}
M.A.~Acero et~al., \emph{{White Paper on Light Sterile Neutrino Searches and
  Related Phenomenology}},  \href{https://arxiv.org/abs/2203.07323}{{\ttfamily
  2203.07323}}.

\bibitem{Adam:1981prem}
A.M.~Dziewonski and D.L.~Anderson, \emph{Preliminary reference earth model},
  {\emph{Physics of the Earth and Planetary Interiors} {\bfseries 25} (1981)
  297}.

\bibitem{Honda:2006qj}
M.~Honda, T.~Kajita, K.~Kasahara, S.~Midorikawa and T.~Sanuki,
  \emph{{Calculation of atmospheric neutrino flux using the interaction model
  calibrated with atmospheric muon data}},
  \href{https://doi.org/10.1103/PhysRevD.75.043006}{\emph{Phys. Rev. D}
  {\bfseries 75} (2007) 043006}
  [\href{https://arxiv.org/abs/astro-ph/0611418}{{\ttfamily
  astro-ph/0611418}}].

\bibitem{Sanuki:2006yd}
T.~Sanuki, M.~Honda, T.~Kajita, K.~Kasahara and S.~Midorikawa, \emph{{Study of
  cosmic ray interaction model based on atmospheric muons for the neutrino flux
  calculation}}, \href{https://doi.org/10.1103/PhysRevD.75.043005}{\emph{Phys.
  Rev. D} {\bfseries 75} (2007) 043005}
  [\href{https://arxiv.org/abs/astro-ph/0611201}{{\ttfamily
  astro-ph/0611201}}].

\bibitem{Gaisser:2013bla}
T.K.~Gaisser, T.~Stanev and S.~Tilav, \emph{{Cosmic Ray Energy Spectrum from
  Measurements of Air Showers}},
  \href{https://doi.org/10.1007/s11467-013-0319-7}{\emph{Front. Phys.
  (Beijing)} {\bfseries 8} (2013) 748}
  [\href{https://arxiv.org/abs/1303.3565}{{\ttfamily 1303.3565}}].

\bibitem{Hoerandel:2002yg}
J.R.~Hoerandel, \emph{{On the knee in the energy spectrum of cosmic rays}},
  \href{https://doi.org/10.1016/S0927-6505(02)00198-6}{\emph{Astropart. Phys.}
  {\bfseries 19} (2003) 193}
  [\href{https://arxiv.org/abs/astro-ph/0210453}{{\ttfamily
  astro-ph/0210453}}].

\bibitem{Zatsepin:2006ci}
V.I.~Zatsepin and N.V.~Sokolskaya, \emph{{Three component model of cosmic ray
  spectra from 100-gev up to 100-pev}},
  \href{https://doi.org/10.1051/0004-6361:20065108}{\emph{Astron. Astrophys.}
  {\bfseries 458} (2006) 1}
  [\href{https://arxiv.org/abs/astro-ph/0601475}{{\ttfamily
  astro-ph/0601475}}].

\bibitem{Ostapchenko:2010vb}
S.~Ostapchenko, \emph{{Monte Carlo treatment of hadronic interactions in
  enhanced Pomeron scheme: I. QGSJET-II model}},
  \href{https://doi.org/10.1103/PhysRevD.83.014018}{\emph{Phys. Rev. D}
  {\bfseries 83} (2011) 014018}
  [\href{https://arxiv.org/abs/1010.1869}{{\ttfamily 1010.1869}}].

\bibitem{Riehn:2015oba}
F.~Riehn, R.~Engel, A.~Fedynitch, T.K.~Gaisser and T.~Stanev, \emph{{A new
  version of the event generator Sibyll}},
  \href{https://doi.org/10.22323/1.236.0558}{\emph{PoS} {\bfseries ICRC2015}
  (2016) 558} [\href{https://arxiv.org/abs/1510.00568}{{\ttfamily
  1510.00568}}].

\bibitem{IceCube-Gen2:2020qha}
{\scshape IceCube-Gen2} collaboration, \emph{{IceCube-Gen2: the window to the
  extreme Universe}}, \href{https://doi.org/10.1088/1361-6471/abbd48}{\emph{J.
  Phys. G} {\bfseries 48} (2021) 060501}
  [\href{https://arxiv.org/abs/2008.04323}{{\ttfamily 2008.04323}}].

\bibitem{IceCube:2024kel}
{\scshape IceCube} collaboration, \emph{{A search for an eV-scale sterile
  neutrino using improved high-energy $\nu_\mu$ event reconstruction in
  IceCube}},  \href{https://arxiv.org/abs/2405.08070}{{\ttfamily 2405.08070}}.

\bibitem{Esmaili:2014gya}
A.~Esmaili, O.L.G.~Peres and P.D.~Serpico, \emph{{Impact of sterile neutrinos
  on the early time flux from a galactic supernova}},
  \href{https://doi.org/10.1103/PhysRevD.90.033013}{\emph{Phys. Rev. D}
  {\bfseries 90} (2014) 033013}
  [\href{https://arxiv.org/abs/1402.1453}{{\ttfamily 1402.1453}}].

\bibitem{Ohlsson:1999xb}
T.~Ohlsson and H.~Snellman, \emph{{Three flavor neutrino oscillations in
  matter}}, \href{https://doi.org/10.1063/1.533270}{\emph{J. Math. Phys.}
  {\bfseries 41} (2000) 2768}
  [\href{https://arxiv.org/abs/hep-ph/9910546}{{\ttfamily hep-ph/9910546}}].

\end{thebibliography}\endgroup

\appendix

\section{\label{sec:Cayley_32}Cayley-Hamilton formalism in (3+2) scenario}

In this appendix we provide the formulae for the computation of oscillation probabilities in $(3+2)$ scenario using the Cayley-Hamilton formalism for $5\times5$ matrices, generalizing the formulae for $(3+1)$ scenario in~\cite{Esmaili:2014gya}. When neutrinos propagate a distance $r=L$ in the medium, a solution of Eq.~\eqref{ODE} for constant matter density can be written in terms of the evolution operator $S(L) = e^{-i\mathcal{H}L}$. The exponential of $\mathcal{H}$ can be written as a polynomial of the matrix $T = \mathcal{H}-(\text{tr} \mathcal{H})I/5$, which is the traceless part of the Hamiltonian and therefore its eigenvalues $\lambda_{i}$ satisfy the following condition:
\begin{equation}
\lambda_{1}+\lambda_{2}+\lambda_{3}+\lambda_{4}+ \lambda_{5}= 0~.
\label{con_Tdiag}
\end{equation}

The general approach to find the exponential of an $N\times N$ matrix with the Cayley-Hamilton formalism is described in the section~III of Ref.~\cite{Ohlsson:1999xb}. The evolution operator, $S(L)$, can be written as 
\begin{equation}
S(L) = \phi\, e^{-iTL} = \phi\, \sum_{k=0}^{4} a_{k}(-iLT)^{k}~,
\label{operator} 
\end{equation}
where $\phi = e^{-i(\text{tr}\mathcal{H})L/5}$ is a phase factor and $a_{k} \; (k=0,\dots, 4)$ are some coefficients that will be computed in the following. Using the eigenvalues $\lambda_{i}$, one can rewrite $e^{-iTL}$ as
\begin{equation}
    e^{-i\lambda_{i}L}=\sum_{k=0}^{4} a_{k}(-iL\lambda_{i})^{k}~, \quad i=1,\dots , 5~.
\label{coeff}
\end{equation}
Thus, the coefficients $a_{k}$ can be determined from the previous equation and its substitution into Eq.~\eqref{operator}. After factorizing the terms proportional to $e^{-i\lambda_{i}L}$, and using Eq.~\eqref{con_Tdiag} to simplify the coefficient of $T^{3}$, we obtain
\begin{align*}
    S(L) \; = & \; \frac{e^{-i\lambda_{1}L}}{(\lambda_{2}-\lambda_{1})(\lambda_{3}-\lambda_{1})(\lambda_{4}-\lambda_{1})(\lambda_{5}-\lambda_{1})} \; [\lambda_{2}\lambda_{3}\lambda_{4}\lambda_{5}I - (\lambda_{2}\lambda_{3}\lambda_{4} + \lambda_{2}\lambda_{3}\lambda_{5}+\lambda_{2}\lambda_{4}\lambda_{5}+\lambda_{3}\lambda_{4}\lambda_{5})T +\\
    & (\lambda_{2}\lambda_{3}+\lambda_{2}\lambda_{4}+\lambda_{2}\lambda_{5}+\lambda_{3}\lambda_{4}+\lambda_{3}\lambda_{5}+\lambda_{4}\lambda_{5})T^{2} +\lambda_{1}T^{3} + T^{4}]+ \\
    & \frac{e^{-i\lambda_{2}L}}{(\lambda_{1}-\lambda_{2})(\lambda_{3}-\lambda_{2})(\lambda_{4}-\lambda_{2})(\lambda_{5}-\lambda_{2})} \; [\lambda_{1}\lambda_{3}\lambda_{4}\lambda_{5}I - (\lambda_{1}\lambda_{3}\lambda_{4} + \lambda_{1}\lambda_{3}\lambda_{5}+\lambda_{1}\lambda_{4}\lambda_{5}+\lambda_{3}\lambda_{4}\lambda_{5})T+\\
    & (\lambda_{1}\lambda_{3}+\lambda_{1}\lambda_{4}+\lambda_{1}\lambda_{5}+\lambda_{3}\lambda_{4}+\lambda_{3}\lambda_{5}+\lambda_{4}\lambda_{5})T^{2} + \lambda_{2}T^{3} + T^{4}]+ \\
    & \frac{e^{-i\lambda_{3}L}}{(\lambda_{1}-\lambda_{3})(\lambda_{2}-\lambda_{3})(\lambda_{4}-\lambda_{3})(\lambda_{5}-\lambda_{3})} \; [\lambda_{1}\lambda_{2}\lambda_{4}\lambda_{5}I - (\lambda_{1}\lambda_{4}\lambda_{5} + \lambda_{2}\lambda_{4}\lambda_{5}+\lambda_{1}\lambda_{2}\lambda_{4}+\lambda_{1}\lambda_{2}\lambda_{5})T+\\
    & (\lambda_{4}\lambda_{5}+\lambda_{2}\lambda_{4}+\lambda_{2}\lambda_{5}+\lambda_{1}\lambda_{2}+\lambda_{1}\lambda_{4}+\lambda_{1}\lambda_{5})T^{2} +\lambda_{3}T^{3} + T^{4}]+ \\
    & \frac{e^{-i\lambda_{4}L}}{(\lambda_{1}-\lambda_{4})(\lambda_{2}-\lambda_{4})(\lambda_{3}-\lambda_{4})(\lambda_{5}-\lambda_{4})} \; [\lambda_{1}\lambda_{2}\lambda_{3}\lambda_{5}I - (\lambda_{1}\lambda_{3}\lambda_{5} + \lambda_{2}\lambda_{3}\lambda_{5}+\lambda_{1}\lambda_{2}\lambda_{3}+\lambda_{1}\lambda_{2}\lambda_{5})T+\\
    & (\lambda_{3}\lambda_{5}+\lambda_{2}\lambda_{3}+\lambda_{2}\lambda_{5}+\lambda_{1}\lambda_{2}+\lambda_{1}\lambda_{3}+\lambda_{1}\lambda_{5})T^{2} +\lambda_{4}T^{3} + T^{4}]+ \\
    & \frac{e^{-i\lambda_{5}L}}{(\lambda_{1}-\lambda_{5})(\lambda_{2}-\lambda_{5})(\lambda_{3}-\lambda_{5})(\lambda_{4}-\lambda_{5})} \; [\lambda_{1}\lambda_{2}\lambda_{3}\lambda_{4}I - (\lambda_{1}\lambda_{3}\lambda_{4} + \lambda_{2}\lambda_{3}\lambda_{4}+\lambda_{1}\lambda_{2}\lambda_{3}+\lambda_{1}\lambda_{2}\lambda_{4})T+\\
    & (\lambda_{3}\lambda_{4}+\lambda_{2}\lambda_{3}+\lambda_{2}\lambda_{4}+\lambda_{1}\lambda_{2}+\lambda_{1}\lambda_{3}+\lambda_{1}\lambda_{4})T^{2} +\lambda_{5}T^{3} + T^{4}]~,
\end{align*}
where $I$ is the $5\times5$ identity matrix. After some algebra, simplifying the coefficients, we obtain
\begin{align}
\nonumber S(L)= \phi \sum_{i=1}^{5} &\frac{e^{-i\lambda_{i}L}}{c_{1}+2c_{2}\lambda_{i}+3c_{3}\lambda_{i}^{2}+5\lambda_{i}^{4}}\; [(c_{1}+c_{2}\lambda_{i}+c_{3}\lambda_{i}^{2}+\lambda_{i}^{4})I+\\
    & (c_{2}+c_{3}\lambda_{i}+\lambda_{i}^{3})T+(c_{3}+\lambda_{i}^{2})T^{2}+\lambda_{i}T^{3}+T^{4}]~,
    \label{Sfor5nu}
\end{align}
where $\lambda_{i}$'s are the roots of the characteristic equation 
\begin{equation*}
    \lambda^{5}+c_{3}\lambda^{3}+c_{2}\lambda^{2} + c_{1}\lambda + c_{0}=0~,
    \label{characteristic_eq}
\end{equation*}
with the coefficients $c_{a}$ given by
\begin{align*}
 c_{0} = & \; \text{det}(T) = \lambda_{1}\lambda_{2}\lambda_{3}\lambda_{4}\lambda_{5} = 0~,\\
 c_{1} = & \; -\text{tr}(T^{4})/4 \\ 
 = & \;  \lambda_{1}\lambda_{2}\lambda_{3}\lambda_{4}+\lambda_{1}\lambda_{2}\lambda_{3}\lambda_{5}+\lambda_{1}\lambda_{3}\lambda_{4}\lambda_{5}+\lambda_{2}\lambda_{3}\lambda_{4}\lambda_{5}+\lambda_{1}\lambda_{2}\lambda_{4}\lambda_{5}~,\\   
 c_{2} = & \; -\text{tr}(T^{3})/3 \\
 = & \; -(\lambda_{1}\lambda_{2}\lambda_{3}+\lambda_{1}\lambda_{2}\lambda_{4}+\lambda_{1}\lambda_{2}\lambda_{5}+\lambda_{1}\lambda_{4}\lambda_{5} +\lambda_{1}\lambda_{3}\lambda_{5}+\lambda_{2}\lambda_{3}\lambda_{4}\\
 & +\lambda_{2}\lambda_{3}\lambda_{5}+\lambda_{3}\lambda_{4}\lambda_{5}+\lambda_{1}\lambda_{3}\lambda_{4}+\lambda_{2}\lambda_{4}\lambda_{5})~,\\
 c_{3} = & \; -\text{tr}(T^{2})/2 \\ 
 = & \; \lambda_{1}\lambda_{2}+\lambda_{1}\lambda_{3}+\lambda_{1}\lambda_{4}+\lambda_{1}\lambda_{5}+\lambda_{2}\lambda_{3}+\lambda_{2}\lambda_{4}+\lambda_{2}\lambda_{5}\\
 & +\lambda_{3}\lambda_{4}+\lambda_{3}\lambda_{5}+\lambda_{4}\lambda_{5}~.
\end{align*}

For the neutrino propagation in a medium with variable density, one can approximate the electron number density profile by the number of $k$ layers with constant density. Then, the total evolution operator is given by 
\begin{equation*}
S = S_{k} S_{k-1} \cdots S_{2} S_{1}~,
\end{equation*}
where $S_i$ is the evolution operator of layer $i$. Once the evolution operator is calculated, the oscillation probability the neutrino flavor $\alpha$ to flavor $\beta$ at a distance $L$, can be computed by $P(\nu_{\alpha} \rightarrow \nu_{\beta}; L) \equiv P_{\alpha \beta}(L)=\vert \mathit{S}_{\alpha \beta}(L)\vert^{2}$.\\

\end{document}